\documentclass[twocolumn,prb,amsmath,amssymb,floatfix,superscriptaddress]{revtex4-2} 
\usepackage{amsmath}
\usepackage{amssymb}
\usepackage{graphicx}
\usepackage{bm}
\usepackage{color}
\usepackage{hyperref}
\usepackage{graphicx}
\usepackage{upgreek}
\usepackage{scalerel}
\usepackage{multirow}
 
 \usepackage[bb=dsserif]{mathalpha}
 
\usepackage{pgffor}
\usepackage{pdfpages}
\usepackage{mathdots}
\usepackage{float}
\usepackage{silence}
\usepackage{physics}    
\usepackage{lscape}   
\usepackage{color, colortbl}
\usepackage[table]{xcolor}
\usepackage{comment}  
\usepackage{soul}
\usepackage{ulem}

\makeatletter
\AtBeginDocument{\let\LS@rot\@undefined}
\makeatother

\begin{document}
\title{Odd-frequency superconducting pairing due to multiple Majorana edge modes in driven topological superconductors}
\author{Eslam Ahmed}
\affiliation{Department of Applied Physics, Nagoya University, Nagoya 464-8603, Japan}
\author{Shun Tamura}
\affiliation{Institute for Theoretical Physics and Astrophysics, University of W\"{u}rzburg, D-97074 W\"{u}rzburg, Germany}
\author{Yukio Tanaka}
\affiliation{Department of Applied Physics, Nagoya University, Nagoya 464--8603, Japan}
\affiliation{Research Center for Crystalline Materials Engineering, Nagoya University, Nagoya 464-8603, Japan}
\author{Jorge Cayao}
\affiliation{Department of Physics and Astronomy, Uppsala University, Box 516, S-751 20 Uppsala, Sweden}
\date{\today} 
\begin{abstract}
Majorana zero modes have been shown to be the simplest quasiparticles exhibiting pure odd-frequency pairing, an effect that has  so far been theoretically established in the static regime. In this work, we investigate the formation of Majorana modes and odd-frequency pairing in  $p$-wave spin-polarized superconductors under a time-dependent drive. We first show  that the driven system hosts multiple Majorana modes emerging at zero and $\pi$, whose formation can be  controlled by an appropriate tuning of the drive frequency and chemical potential, in agreement with previous studies. Then we explore the induced pair correlations and find that odd-frequency spin-polarized $s$-wave pairing is broadly induced, acquiring large values in the presence of Majorana modes. We discover that, while odd-frequency pairing is proportional to $\sim1/\omega$ in the presence of Majorana zero modes,  it is proportional to $\sim 1/(\omega-\pi\hbar/T)$  in the presence of Majorana $\pi$ modes,  where $T$ is the periodicity of the  drive.   Furthermore, we find that the amount of odd-frequency pairing becomes larger when multiple Majorana modes appear but the overall divergent profile  as a function of frequency remains.  We also show that the divergent odd-frequency pairing   is robust against scalar disorder. Notably, we establish a bulk-boundary correspondence between the amount of boundary odd-$\omega$ pairing and the bulk topological invariants in driven chiral systems, which we show to be  protected by chiral symmetry and is thus robust against disorder. Our work thus paves the way for understanding the emergent pair correlations in driven topological superconductors

\end{abstract}
\maketitle

\section{Introduction}
Topological superconductors are characterized by the emergence of Majorana zero modes (MZMs) \cite{RevModPhys.83.1057,tanaka2011symmetry,leijnse2012introduction,beenakker2013search,sato2017topological,TanakaCayaotheory}, charge neutral and zero energy quasiparticles with potential for realizing topological qubits \cite{sarma2015majorana,Lahtinen_2017,beenakker2019search,aguado2020majorana,aguado2020perspective,Marra_2022}.  While charge neutrality and zero energy signatures have been extensively pursued to detect MZMs \cite{lutchyn2018majorana,zhang2019next,prada2019andreev,frolov2019quest,flensberg2021engineered,Marra_2022}, they do not unambiguously probe the emergence of Majorana physics \cite{PhysRevB.91.024514,PhysRevB.104.134507,prada2019andreev}. Another less explored characteristic of MZMs is that their pair amplitude is an odd function in the relative time, or frequency, revealing  odd-frequency pairing as an intriguing property of MZMs \cite{TanakaTanuma2007PRB,tanaka2011symmetry,tanaka2018surface,cayao2019odd,RevModPhys.91.045005,Tanaka2021,TanakaCayaotheory}. The odd-frequency pairing of MZMs has so far only attracted theoretical studies, which predict a divergent  pair amplitude at zero frequency as an unambiguous signature of Majorana physics \cite{tanaka2018surface,cayao2019odd,RevModPhys.91.045005,Tanaka2021,TanakaCayaotheory}.

The relationship between odd-frequency  pairing and MZMs has been theoretically studied in several systems with topological superconductivity. This relationship has been investigated in  heterostructures based on $p$-wave superconductors \cite{PhysRevB.70.012507,Takagi18,tamura18,thanos2019,PhysRevB.101.094506}, $d$-wave superconductors \cite{PhysRevB.99.184501}, topological insulators \cite{PhysRevB.86.075410,PhysRevB.86.144506,PhysRevB.87.220506,PhysRevB.92.205424,Lu_2015,PhysRevB.92.100507,PhysRevB.96.155426,PhysRevB.96.174509,bo2016,PhysRevB.97.075408,PhysRevLett.120.037701,PhysRevB.97.134523,PhysRevB.101.180512,PhysRevB.106.L100502}, Weyl semimetals \cite{PhysRevB.100.104511,parhizgar2020large},  semiconductors with Rashba spin-orbit coupling \cite{PhysRevB.87.104513,Ebisu16,PhysRevB.98.075425,PhysRevB.99.184501},  Majorana nanowires \cite{PhysRevB.95.184506,PhysRevB.101.094506},  Sachdev-Ye-Kitaev setups \cite{PhysRevB.99.024506}, and interacting MZMs \cite{PhysRevB.92.121404}. All these works helped to establish a strong connection between MZMs and odd-frequency pairing and revealed the exotic superconducting nature of MZMs \cite{golubov2009odd,tanaka2011symmetry,tanaka2018surface,cayao2019odd,RevModPhys.91.045005}. In fact, the oddness in frequency of the superconducting pairing implies that odd-frequency pairing is an effect that is nonlocal in time and, therefore, can be seen as an intrinsic dynamical phenomenon  \cite{RevModPhys.91.045005,balatsky2020quantum}. Despite the dynamical nature of odd-frequency pairing, all previous studies addressed its relationship with MZMs only in   static topological superconductors, leaving  the time-dependent regime unexplored.

 \begin{figure}[!t]
\centering
	\includegraphics[width=0.8\columnwidth]{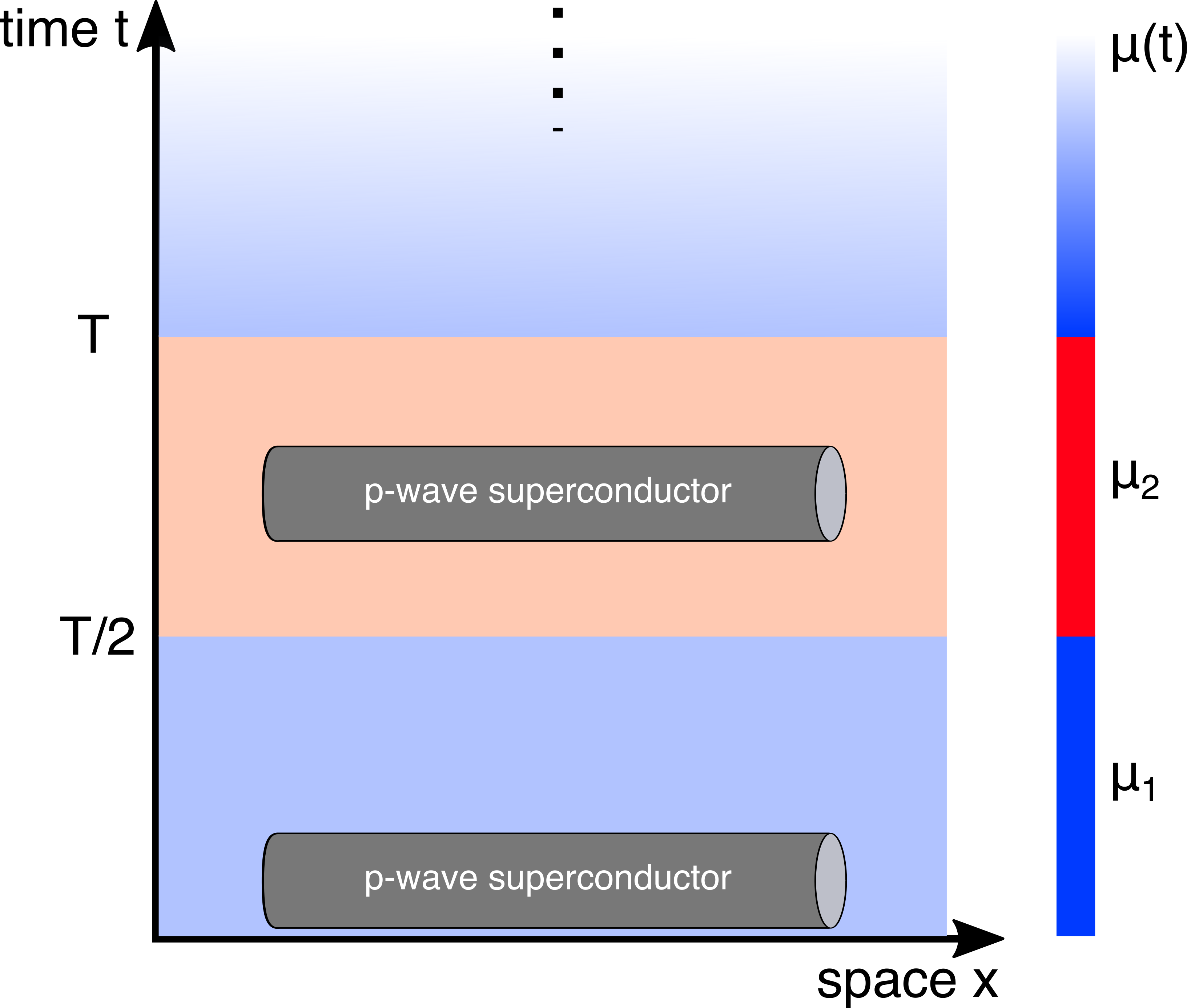}\\
 	\caption{Sketch of a $p$-wave superconductor (gray cylinder) under a time-periodic chemical potential $\mu(t)$ with periodicity $T$. Here,  $\mu(t)$ depends on time in a piece-wise manner, so that the total time-dependent system is described by the same Hamiltonian but with different chemical potentials at every half-cycle, see   blue and red colored regions. }
\label{Fig0} 
\end{figure}

In this work, we consider $p$-wave superconductors driven by a time-periodic field and investigate the emergence of Majorana states and odd-frequency pairing. In particular, we focus on time-periodic modulations in the chemical potential of the superconductor and treat the time-dependent problem within Floquet theory.  Since our system is a  $p$-wave superconductor, it hosts a topological phase with MZMs but, as expected, the drive also induces a Floquet topological phase with Majorana modes at quasienergies equal to $\pi\hbar/T$,  known as Majorana $\pi$ modes (MPMs), where $T$ is the period of the drive \cite{PhysRevLett.106.220402}. We also show that the driven $p$-wave superconductor  can simultaneously host multiple MZMs and MPMs, an effect that is supported by higher values of their topological invariants and can be fully controlled by the period of the drive   and chemical potentials;  this is consistent with previous works on other time-periodic topological superconductors \cite{sym14122546,PhysRevB.107.035427,PhysRevB.108.L081403,PhysRevB.101.014306,PhysRevB.88.155133}.   Notably, we discover that odd-frequency spin-polarized $s$-wave pairs are induced and enhanced in the presence of MZMs and MPMs,  acquiring a unique behaviour proportional to $\sim1/\omega$ and $\sim 1/(\omega-\pi\hbar/T)$, respectively. This, therefore, suggests a strong and intriguing relationship between dynamical topological superconductivity and  odd-frequency pairing.  We  then demonstrate that this relationship stems from the spectral bulk-boundary correspondence, which relates the amount of odd-$\omega$ pairing and the bulk topological invariant, a result only proven before in the static regime \cite{tamura18,PhysRevB.100.174512}. Furthermore, we show that this relationship is protected by chiral symmetry, which explains its robustness against scalar disorder and highlights the stability of the emergent odd-frequency pairing  in driven topological superconductors.  Our results provide a new way to generate and manipulate odd-frequency pairing using Floquet engineering and offer fundamental understanding of the emerging pair correlations in driven topological superconductors.  
 
 The remainder of this article is organized as follows. In  Section \ref{section2} we introduce the  time-periodic $p$-wave superconductor model  and discuss the Floquet method. In Section \ref{section3} we study the emergence of MZMs and MPMs obtained within a Floquet description. In Section \ref{section4} we apply the Floquet method to study the pair amplitudes and obtain the odd-frequency pairing  and also test its robustness against scalar disorder.  Finally, in Section \ref{section5}, we present our conclusions. To further support this work, in Appendices \ref{AppendixB}, \ref{AppendixA}, \ref{app:c}, and \ref{app:d} we present additional details of the calculations.
 
\section{Time-periodic topological superconductors}
\label{section2}
We consider a  finite one-dimensional chain of spin-polarized fermions 
with $p$-wave pair potential subjected to time-periodic modulations in the chemical potential, which is modeled by 
\begin{equation}
\label{eq1}
\begin{split}
    H(t) &= \sum_{j=1}^N \psi_j^\dagger(-\mu(t)\tau_z)\psi_j\\
    &+  \sum_{j=1}^{N-1}\psi_j^\dagger(-w\tau_z+i\Delta\tau_y)\psi_{j+1}+{\rm h.c.}\,,
    \end{split}
\end{equation}
where $\psi_j=(c_j,c^\dagger_j)^\mathrm{T}$ is the Nambu spinor at site $j$, $c_j$ ($c_j^{\dagger}$) destroys (creates) an electronic state at site $j$, $\tau_{j}$ is the $j$-th Pauli matrix in Nambu space, $\Delta$ is the $p$-wave order parameter, $w$ is the nearest-neighbour hopping amplitude,  and $\mu(t)$ represents the time-periodic chemical potential. The length of the system is given by L = Na, where N represents the number of sites and a=1 is the lattice spacing. Here, we consider a driving protocol for the chemical potential $\mu(t)$ such that it is given by piece-wise time-periodic modulations as
\begin{equation}
\label{eqmut}
        \mu(t) = \begin{cases} 
                \mu_1,  & nT < t \leq (n+\frac{1}{2})T\,,
                    \\
                \mu_2, & (n+\frac{1}{2})T < t \leq (n+1)T\,,
            \end{cases}
\end{equation}
where  $T$ is the driving period and $n\in \mathbb{Z}$. Therefore,   the total time-dependent Hamiltonian   governing the driven   system is determined by two piece-wise constant Hamiltonians as a function of time, denoted here as $H_{1,2}$, such that
\begin{equation}\label{eq:Hamiltonian}
    H(t) = \begin{cases}
            H_1, \qquad  nT < t \leq (\frac{1}{2}+n)T\,,
            \\
            H_2,  \qquad  (\frac{1}{2} + n)T < t \leq (n+1)T\,.
            \end{cases}
\end{equation}
The only difference between $H_{1,2}$ is that they
have distinct chemical potentials at every half-cycle given by Eq.\,(\ref{eqmut}). In the static regime, when there is no time dependence in the chemical potential, such that $\mu_{1,2}\equiv \mu$, the $p$-wave superconductor model given by Eq.(\ref{eq1}) describes the well-known   Kitaev   chain and has been shown to host a topological phase with MZMs when $\mu<|2w|$, see e.g.,\,Refs.\,\cite{leijnse2012introduction,sato2017topological}. MZMs emerge as edge states whose wavefunctions exponentially decay towards the bulk of the superconductor and their energies reach zero for sufficiently large systems.  Recent experiments in semiconductor-superconductor hybrids have shown that realizing topological superconductivity based on the Kitaev chain is within experimental reach \cite{zhang2019next,prada2019andreev,frolov2019quest,flensberg2021engineered}.

Motivated by the exotic properties of the Kitaev chain, here we are interested in exploring its behaviour and how its topological properties  vary by  applying a time-periodic modulation in the chemical potential. Specifically, we are interested  in engineering and controlling the emergent topological superconductivity, its Majorana modes, and the superconducting pair correlations by means of the time-periodic drive. For this purpose, we  focus on the  stroboscopic evolution of the system  which is defined at integer multiple of the driving period $T$ \cite{Floquet.quantum.systems.engineering}. At stroboscopic times, the evolution is governed by the time-evolution operator over one period $U_T[t_0]\equiv U(t_{0}+T,t_{0})$, with
\begin{equation}
\label{UTt0}
U_T[t_0] = \mathcal{T} \exp\left(-\frac{i}{\hbar}\int_{t_0}^{t_0+T}H(s)\dd s\right)\,,
\end{equation}
where $\mathcal{T}$ is the time-ordering operator, $t_{0}$ is the initial time, and $H$ is described by Eq.\,(\ref{eq1}). We note that $U_T[t_0]$ is time-periodic, namely, $U_T[t_0]\equiv U(t_{0}+T,t_{0})=U(t_{0}+2T,t_{0}+T)$,  and is known as the Floquet time evolution operator.  Then, the effective Hamiltonian for the stroboscopic evolution, at $t_{0}+nT$ for $n\in\mathbb{Z}$, also known as the Floquet Hamiltonian \cite{Floquet.quantum.systems.engineering}, is obtained as
\begin{equation}
\label{Heff}
    H_{\rm F}[t_0] = \frac{i\hbar}{T}\ln(U_T[t_0])\,.
\end{equation}
where ${\ln}$ here represents the natural logarithm.  We note that the effective Hamiltonian $H_{\rm F}$  depends on the choice of the initial time $t_0$, which is a gauge choice and completely arbitrary, often taken  to make $H_{\rm F}$ acquire its simplest form \cite{high.frequency.Floquet.engineering}. Moreover, $H_{\rm F}$  also depends on the  logarithm branch cut, considered in a way  that the eigenvalues of $H_{\rm F}$, also known as \textit{quasienergies}, belong to the region 
$[-{\pi\hbar}/{T},{\pi\hbar}/{T})$  where  $\abs{\ev{H_{\rm F}}{ \psi}}\leq{\pi\hbar}/{T}$ holds for an arbitrary stationary state $\ket{\psi}$ \cite{high.frequency.Floquet.engineering}. We are interested in exploring the formation of Majorana modes and also their impact on the emergence of odd-frequency pair correlations. Below we address these points by using the effective Hamiltonian Eq.\,(\ref{Heff}).


\section{Majorana edge modes: Topological invariants and quasienergy spectrum}
\label{section3}
We start by investigating the emergence of   Majorana edge modes, expected to appear when the driven system described by Eq.\,(\ref{Heff}) becomes topological.  For this purpose, we next identify these topological regimes by calculating the topological invariants and also characterize their Majorana edge modes in the quasienergy spectrum.

\subsection{Bulk quasienergies and multiple Fermi surfaces}
\label{subsection3A}
We first analyze the  quasienergies of the bulk Floquet Hamiltonian given by Eq.\,(\ref{Heff}).  To calculate the bulk quasienergies, we assume periodic boundary conditions and study Eq.\eqref{Heff} in momentum space. Since our Hamiltonian is block-diagonal in momentum space, we can write the Floquet Hamiltonian as
\begin{equation}
    H_{\rm F}[t_0] = \sum_k \psi_k^\dagger H_{\rm F}(k)[t_0]\psi_k\,,
\end{equation}
where $\psi_k=\frac{1}{\sqrt{N}}\sum_{j=1}^Ne^{-ikj}\psi_j$ is the Fourier transform of $\psi_j$ and the bulk Floquet Hamiltonian $H_{\rm F}(k)[t_0]$ is given by
\begin{equation}
    e^{-i\frac{T}{\hbar}H_{\rm F}(k)[t_0]} = \mathcal{T}e^{-i\frac{1}{\hbar}\int_{t_0}^{t_0+T}H(t,k)\dd t}\,,
\end{equation}
with $H(t,k)$ being the momentum representation of the time-dependent Hamiltonian of Eq.\,\eqref{eq1}.  For small period $T$, we have performed  Baker-Campbell-Hausdorff expansion of the Floquet Hamiltonian $H_{\rm F}(k)[t_0]$ in Appendix \ref{AppendixA}. We find that the positive bulk Floquet quasienergy is given by
\begin{equation}
\label{quasienergy}
\begin{split}
    E_{\rm F}(k) &= \frac{\hbar}{T}\arccos \bigg[\cos(E_1T/2\hbar)\cos(E_2T/2\hbar) \\ 
    &- \frac{\bm{E}_1\cdot\bm{E}_2}{E_1E_2}\sin(E_1T/2\hbar)\sin(E_2T/2\hbar) \bigg]\,,
    \end{split}
\end{equation}
while   the negative  band is   given by $-E_{\rm F}$, and $E_j = \abs{\bm{E}_j}$, with $\bm{E}_j= (0,-2\Delta\sin{k},-\mu_j-2w\cos{k})^T$. We refer to     Appendix \ref{AppendixB} for details on the derivation of Eq.\,(\ref{quasienergy}), see also Ref. \cite{PhysRevB.93.184306}. As we see, the bulk quasienergies are symmetric with respect to zero quasienergy due to particle-hole symmetry, and they are also symmetric with respect to quasienergies $\pm\hbar\pi/T$ because they are defined modulus $2\pi$ in Eq.\,(\ref{quasienergy}).  

\begin{figure}[t]
    \centering
    \includegraphics[width=8.6cm]{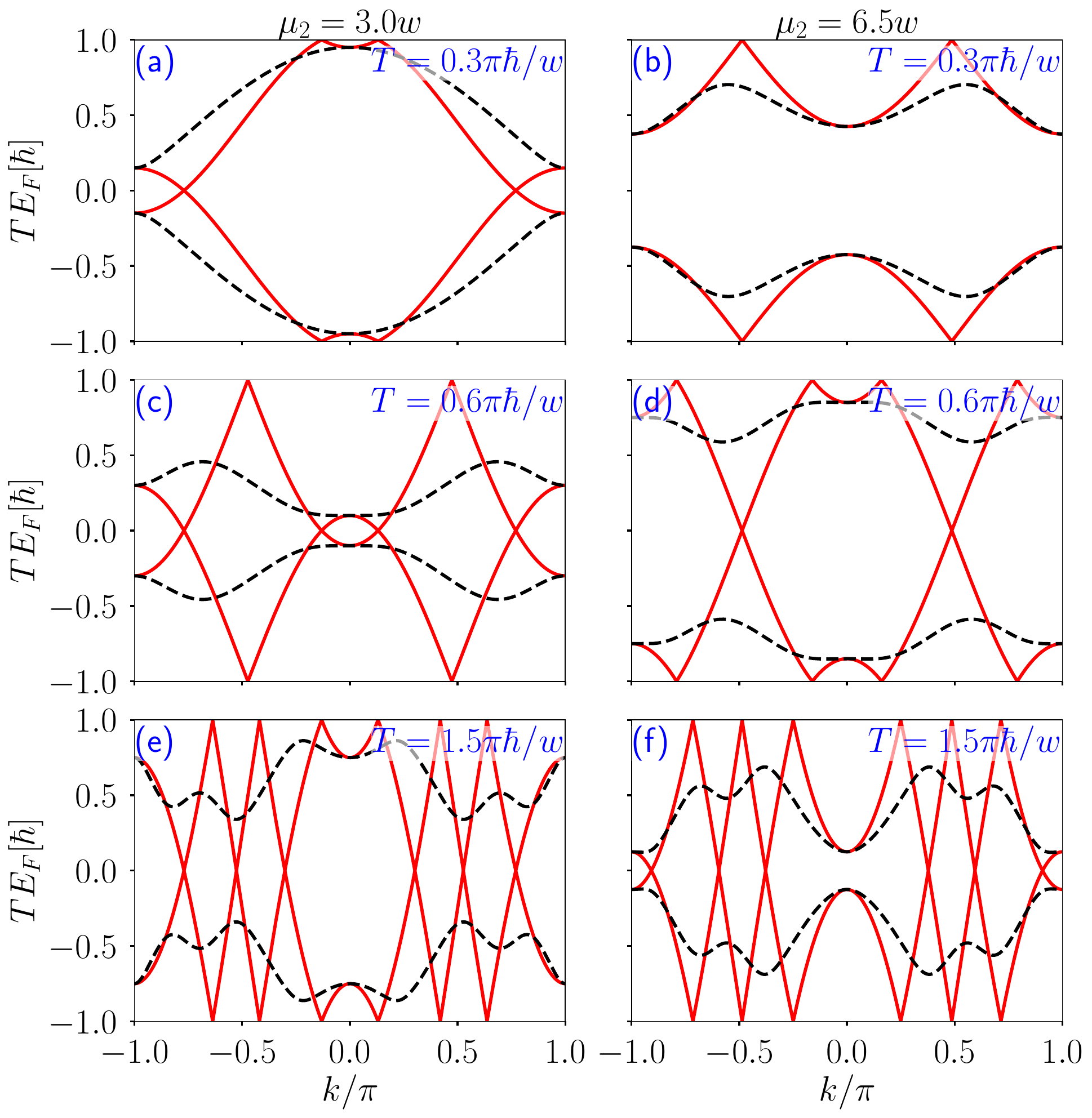}
    \caption{Bulk quasienergy bands  as a function of momentum $k$ for  $\mu_2/w=3$ (left column) and  $\mu_2/w=6.5$ (right column) at $\mu_{1}=0$ and distinct values of the period $T$. Solid red and black dashed    curves depict  the quasienergies at $\Delta=0$ and $\Delta=w$. Top (a,b) , middle (c,d) , and bottom  panels (e,f) correspond to $T=0.3\pi\hbar/w$, $0.6\pi\hbar/w$, and $1.5\pi\hbar/w$, respectively.}
    \label{fig:app_dispersion}
\end{figure}

To visualize $E_\mathrm{\rm F}(k)$, we show its momentum dependence in  Fig.~\ref{fig:app_dispersion} for distinct values of $T$ and  $\mu_{2}$, with $\mu_{1}=0$. For clarity and completeness, we show the bulk quasienergies at zero and finite pair potential, which correspond to $\Delta=0$ and $\Delta\neq0$,   depicted by solid red and dashed black curves, respectively.  The choice of the chemical potentials is motivated by the fact that it allows us to   explore the trivial and topological phases of the Kitaev chain, while the effect of $T$ permits us to inspect the response of the system to the driving field. In the  case with $T\rightarrow0$, the average chemical potential  $(\mu_{1}+\mu_{2})/2$ sets the topological transition, see Appendix \ref{AppendixA}: thus, the regime with $\mu_{2}=3w$ and $\mu_{1}=0$ puts the system in the topological phase, while $\mu_{2}=6.5w$ and $\mu_{1}=0$ corresponds to the trivial phase.

In the normal state, $\Delta=0$, there are multiple Fermi surfaces, which are seen by noting the   crossings at quasienergies $E_{\rm F}=0$ and $E_{\rm F}=\pi\hbar/T$ in  the solid red curves in Fig.~\ref{fig:app_dispersion}. At small $T$ in Fig.~\ref{fig:app_dispersion}(a,b), a Fermi surface appears in the normal state $\Delta=0$ at zero quasienergy $E_{\rm F}=0$ for $\mu_{2}=3w$ but none for $\mu_{2}=6.5w$; intriguingly,   Fermi surfaces appear at $E_{\rm F}=\pi\hbar/T$, revealing an important  effect of the drive. By increasing $T$, multiple Fermi surfaces appear at $E_{\rm F}=0$ and $E_{\rm F}=\pi\hbar/T$ for both cases of  $\mu_{2}$, as seen in Fig.~\ref{fig:app_dispersion}(c-f). Interestingly, the multiple Fermi surfaces only come from the  {\rm log} operation since $H_1$ and $H_2$ commute with each other for $\Delta=0$. This effect is  seen by performing a Baker-Campbell-Hausdorff (BCH) expansion  of the Floquet Hamiltonian in $T$,  where only the lowest order remains and higher orders vanish at $\Delta=0$, see Appendix \ref{AppendixA}. At finite pair potential, we see gap openings at the Fermi surfaces (dashed black curves). At this point, it is worth noting that having a Fermi surface is known to be important for inducing edge  states. In this regard, systems with multiple Fermi surfaces, as it is in our case, are expected to host multiple edge modes, with the number of states related to the number of Fermi surfaces \cite{TanakaCayaotheory}. In our system, the multiple Fermi surfaces  are induced by $T$ and occur at two particular quasienergies, at $E_{\rm F}=0$ and $E_{\rm F}=\pi\hbar/T$, which suggests that our system can be engineered to host multiple edge states under the presence of the drive. To know the number of edge states, however, it is important to go beyond the Fermi surfaces and explore the topological invariants which we carry out in the next subsection.

\subsection{Topological Invariants}
\label{section3a}
To understand the emergence of topological phases in the driven system described by Eq.\,(\ref{Heff}), we   obtain the topological invariants. In this case, it is required to identify the symmetry of the effective Floquet Hamiltonian in Eq.\,(\ref{Heff}).  While  $H(t)$ anti-commutes with the chiral operator $\Gamma = \tau_x$, for chiral symmetry to be preserved in the Floquet Hamiltonian, we need to find an initial time, $t_0$, that ensures a symmetric time evolution.  Specifically, it should satisfy the condition $U_{T}[t_0] = \Gamma A^\dagger \Gamma A$ for a unitary operator $A$. In our case, chiral symmetry is preserved at $t_{0}=\pm T/4$, which then implies that the operator $A$ is given by 
 $A=e^{-\frac{iT}{4\hbar}H_2}e^{-\frac{iT}{4\hbar}H_1}$ for $t_0 = + T/4$;    for $-T/4$  the operator $A$ acquires a similar form but with $H_1$ and $H_2$ exchanged. We  note that  $A$ can  also be written as $U(t_0+T/2,t_0)$, which is the time-evolution operator for a half-cycle and represents an evolution from  $t_0$ to some time $t$ such that the Hamiltonian is time-symmetric around $t$. For our specific system, where chiral symmetry  is preserved at $t_0 = \pm T/4$,   the one-period time-evolution operator is expressed as
\begin{equation}
\label{HFpm}
\begin{split}
U_T^+ &\equiv U_T[T/4] = e^{-\frac{iT}{4\hbar}H_1}e^{-\frac{iT}{2\hbar}H_2}e^{-\frac{iT}{4\hbar}H_1}\,, \\
U_T^- &\equiv U_T[-T/4] = e^{-\frac{iT}{4\hbar}H_2}e^{-\frac{iT}{2\hbar}H_1}e^{-\frac{iT}{4\hbar}H_2}\,,
\end{split}
\end{equation}
where $H_{1,2}$ are given by Eqs.\,(\ref{eq:Hamiltonian}). Then, in the same spirit as for the definition of the Floquet Hamiltonian $H_{\rm F}$ in Eq.\,(\ref{Heff}), here we take Eqs.\,(\ref{HFpm}) and define two chiral-invariant Floquet Hamiltonians denoted as $H_{\rm F}^{\pm}$, namely, $  H_{\rm F}^{\pm} = (i\hbar/T)\ln(U_T^{\pm})$. Thus, using these Hamiltonians, we can define two winding numbers as follows:
\begin{equation}
\label{Wpm}
W^\pm = \frac{1}{4\pi i}\int_{-\pi}^\pi \dd k \Tr{{\Gamma}\big[H_{\rm F}^\pm{(k)}\big]^{-1}\partial_k\big[H_{\rm F}^\pm(k)\big]}\,,
\end{equation}
where $H_{\rm F}^\pm(k)$ represents the effective Hamiltonian in the momentum representation. These winding numbers characterize the change in the total number of  pairs of edge modes.  However, they do not provide detailed information about the number of  MZMs  and  MPMs. This can be remedied by defining two extra winding numbers by combining  $W^\pm$ as \cite{PhysRevB.88.121406,ChiralFloquetWindingNumber,Wu_2023}
\begin{equation}
\label{W0pi}
W_0 = \frac{W^+ + W^-}{2} \quad \text{and} \quad W_\pi = \frac{W^+ - W^-}{2}\,,
\end{equation}
where $W_0$ and $W_\pi$ count the number of  pairs of MZMs and MPMs, respectively.    In practice, to obtain  $W_{0,\pi}$, we discretize the momentum space and perform the integrals in   Eqs.\,(\ref{Wpm})  numerically.

 \begin{figure}[!t]
\centering
        \includegraphics[width=\columnwidth]{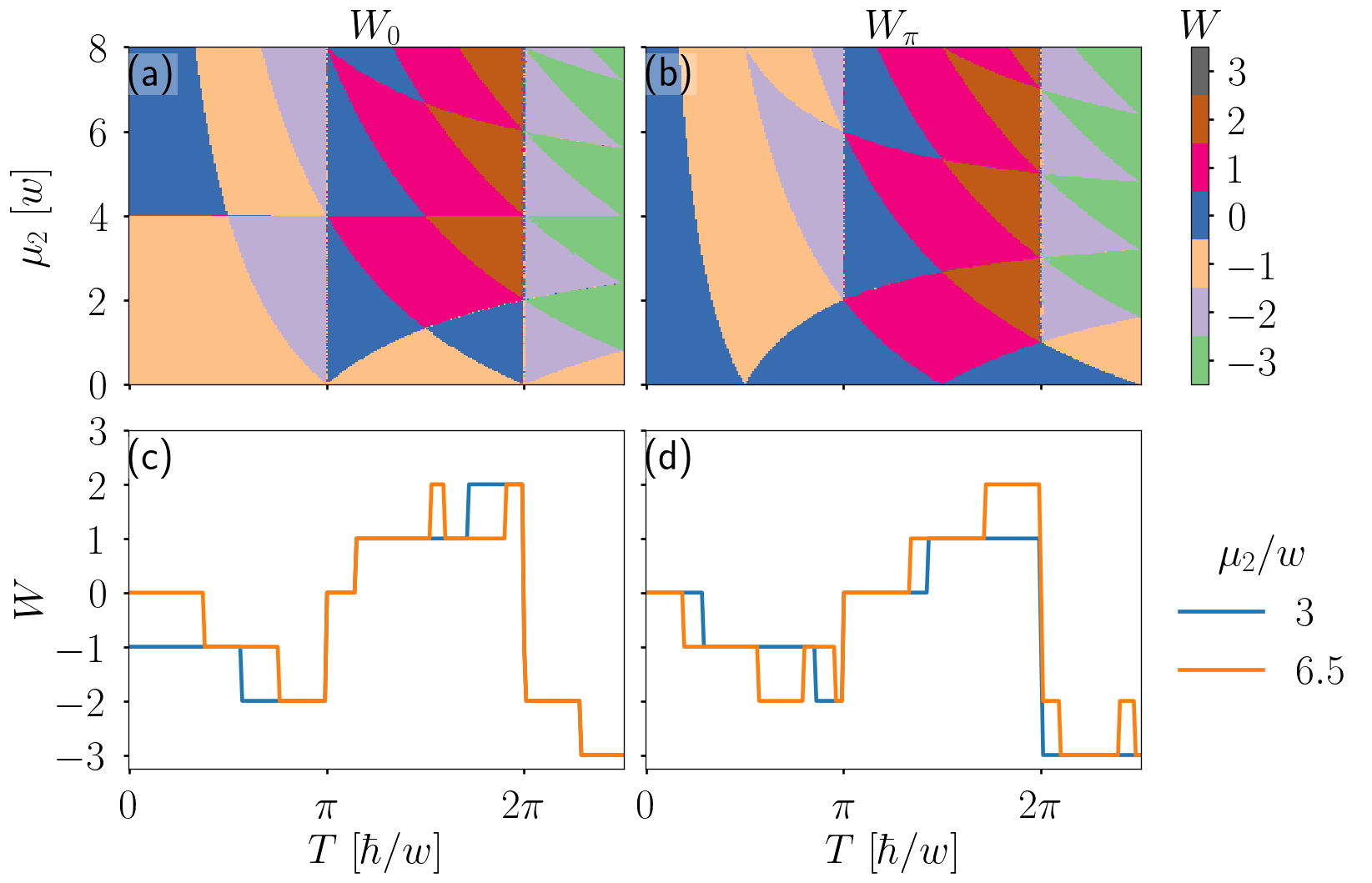}\\
 	\caption{Topological invariants $W_{0}$ (a) and $W_{\pi}$ (b) as a function of the chemical potential $\mu_{2}$ and  period $T$. Panels (c,d) show line cuts of (a,b) as a function of $T$ at fixed $\mu_{2}$. Parameters:   $\Delta=w,\mu_1=0$.}
\label{Fig2} 
\end{figure}

In Fig.\,\ref{Fig2}(a,b) we present the winding numbers $W_{0,\pi}$ as functions of the period $T$ and chemical potential $\mu_{2}$ at fixed $\Delta=w$. We also set to zero  the chemical potential of the first half of the cycle $\mu_1=0$  and introduce a non-zero chemical potential during the second half $\mu_{2}$.   Moreover, in Fig.\,\ref{Fig2}(c,d) we also show line cuts of Fig.\,\ref{Fig2}(a,b)  at fixed $\mu_{2}$. The first observation in  Fig.\,\ref{Fig2}(a,b) is that, near $T=0$, the winding number that counts MZMs takes two values $W_{0}=0,-1$ as a function of $\mu_{2}$, an effect that is consistent with the static $p$-wave superconductor (Kitaev chain) developing a trivial and a topological phase \cite{leijnse2012introduction,sato2017topological}. In fact,  the Floquet Hamiltonian $H_{\rm F}$ becomes $(H_1 + H_2)/2$ near $T=0$, which then implies that the phase transition is dictated by $(\mu_1+\mu_2)/2=2w$, which is consistent with $\mu_{2}=4w$ when $\mu_{1}=0$.  For details on the expansion of $H_{\rm F}$  near $T=0$, see Appendix \ref{AppendixA}. We also note that at $T=0$,  the winding number that counts MPMs is zero ($W_{\pi}=0$) and does not change as $\mu_{2}$ increases, as expected since no MPMs exist in the static regime. 

The behaviour discussed for $W_{0,\pi}$ above is preserved also for very small values of $T$ but undergoes considerable variations as $T$ further increases, revealed in the multicolor regions in Fig.\,\ref{Fig2}(a,b). In fact, we obtain that both $W_{0}$  and $W_{\pi}$ can take integer numbers, with higher values that interestingly predict the emergence of multiple MZMs and also multiple MPMs. 
It is worth noting that the line $\mu_2 = 4w$, which   corresponds to the phase transition criterion for the static Kitaev chain, persists throughout the entire range of $T$. This can be understood by setting $k=\pi$ in Eq.\,\eqref{quasienergy} and observing that the quasienergy given by Eq.\,(\ref{quasienergy}) vanishes $E_{\rm F}=0$ when $\mu_{1}+\mu_{2}=4w$, thus revealing that the topological phase transition of the static system remains robust even in the presence of a time-periodic drive [Fig.\,\ref{Fig2}(a)].  Moreover, from Fig.\,\ref{Fig2}(a,b)  we observe that a notable effect of the drive is its potential to change the topology of the system. For instance, by increasing $T$, a topologically trivial phase can enter into a topologically nontrivial phase with nonzero topological invariant, see blue ($W_{0(\pi)}=0$)-to-magenta ($W_{0(\pi)}=1$) or blue ($W_{0(\pi)}=0$)-to-peach ($W_{0(\pi)}=-1$)  in Fig.\,\ref{Fig2}(a,b).  Furthermore, if the system is topologically nontrivial with a nonzero winding number, by increasing $T$ it is possible to enter into another topological phase with a higher winding number, as seen in the transitions involving magenta ($W_{0(\pi)}=1$)-to-brown ($W_{0(\pi)}=2$) or peach ($W_{0(\pi)}=-1$)-to-violet ($W_{0(\pi)}=-2$) colors, see    Fig.\,\ref{Fig2}(a,b). A simpler visualization of these cases is presented in Fig.\,\ref{Fig2}(c,d) which shows two particular values of $\mu_{2}$ where the system is initially topologically trivial  (orange curve) and nontrivial (blue curve). Therefore, by an appropriate control of the chemical potential $\mu_{2}$ and the drive period $T$, topological phases with multiple Majorana edge modes at zero and at $\pi\hbar/T$ quasienergy can be induced even if the static system is topologically trivial.

\subsection{Quasienergy spectrum}
\label{section3b}
 Having understood the emergence of multiple topological phases in driven $p$-wave superconductors, here we explore their quasienergy spectrum. This is motivated by the fact that MZMs are expected  to appear at zero quasienergy ($E_{\rm F}=0$) while MPMs at $E_{\rm F}=\pi\hbar/T$.  In this regard, we consider
a finite system with $L=100a$ described by Eq.\,(\ref{Heff}) and numerically obtain its eigenvalues. In Fig.\,\ref{Fig3}(a,b) we present the quasienergy spectrum  as a function of $T$ at $\mu_2 =3w$ and $\mu_{2}=6.5w$, respectively. We note that, for visualization purposes, we obtain the spectrum of $T H_{\rm F}$ and that is why the quasienergy gap vanishes   at $T=0$. We fix the chemical potential for the second half of the cycle to these values because it gives us the possibility to explore how the effect of the time-periodic drive impacts the already existent topology and also how it can change the topology of the system, as we discussed in the previous subsection, see Fig.\,\ref{Fig2}(c,d).  The immediate observation is that, in general, the quasienergy spectrum is composed of a dense spectrum which is strongly dependent on $T$ and also of dispersionless quasienergy levels around zero  and $\pi\hbar/T$ quasienergies, which here we refer to as MZMs and MPMs, respectively. 

For $\mu_2 =3w$ in Fig.\,\ref{Fig3}(a), a pair of MZMs appears in the undriven regime, which   is consistent with  the topological invariant $W_{0}$ in Fig.\,\ref{Fig2}(a).   As $T$ increases, the pair of MZMs remains but, interestingly, a pair of MPMs appears at quasienergies of $\pi\hbar/T$, see orange lines in Fig.\,\ref{Fig3}(a). Moreover, at around $T\approx \pi\hbar/(2w)$, the dense spectrum reaches zero quasienergy, giving rise to a gap closing and reopening which then leaves two extra MZMs, and thus the system hosts two pairs of MZMs.  The appearance of multiple MZMs can be seen to be a result of  our driven system having multiple Fermi surfaces, as discussed in Subsection \ref{subsection3A}.  Another possibility is that the considered system exhibits long-range pairing and hopping, an effect that was shown before to lead to multiple Majorana edge modes in the static regime \cite{PhysRevB.88.165111}, see also Refs.\,
\cite{PhysRevB.87.201109,PhysRevB.88.155133,PhysRevB.101.014306,sym14122546,Wu_2023,Soori_2023,PhysRevB.108.L081403,PhysRevB.107.035427,PhysRevB.109.184518,PhysRevB.107.235132}. In Appendix \ref{AppendixA} we show the emergence of long-range  pairing components in the fourth order BCH expansion of $H_{\rm F}$ in $T$, but higher orders are needed for revealing the long-range hopping terms. Something similar happens with the MPMs: at $T\approx \pi\hbar/w$, the dense spectrum touches the quasienergy $\pi\hbar/T$ and leaves two extra MPMs. This behaviour also occurs in Fig.\,\ref{Fig3}(b) but then no MZMs at very low $T$ since here $\mu_{2}=6.5w$ and the static regime is topologically trivial. Nevertheless, the gap closings and emergence of multiple MZMs and MPMs are similar to what occurs in Fig.\,\ref{Fig3}(a) for $\mu_{2}=3w$. Furthermore, we have confirmed that the associated wavefunctions of these multiple Majorana modes are indeed located at the ends of the system, as expected for Majorana wavefunctions due to their spatial nonlocality.  
 
  \begin{figure}[!t]
\centering
	\includegraphics[width=\columnwidth]{Fig4-new.png}\\
 	\caption{(a,b) Quasienergy spectrum as a function of the period $T$ for $\mu_2 =3w$ and $\mu_{2}=6.5w$. Here, MZMs and MPMs are depicted by blue and orange lines, respectively.   Parameters:    $\Delta=w,\mu_1=0$, $N=100$, $\delta/w = 10^{-5}$.}
\label{Fig3} 
\end{figure}
 Thus, dispersionless quasienergy levels depicted in blue and orange lines in  Fig.\,\ref{Fig3}(a,b) represent MZMs and MPMs,  which can be engineered by simply controlling the chemical potential $\mu_{2}$ and the drive period $T$. Since the MZMs and MPMs found here only appear at finite $T$, they reveal a clear impact of the time-periodic drive on   $p$-wave superconductors.

\section{Odd-frequency superconducting pairing at stroboscopic times}
\label{section4}
In this part, we study the nature of the emergent superconducting correlations under the presence of MZMs and MPMs. The superconducting correlations are commonly obtained from the anomalous (electron-hole) part of the system Green's function in Nambu space, see e.g.,\,Refs.\,\cite{zagoskin1998quantum,mahan2013many}. Since the driven system in our case is described by the effective Hamiltonian in Eq.\,(\ref{Heff}) at stroboscopic times, we  can define   an effective Green's function associated with such Hamiltonian, see Appendix \ref{app:d}. Thus, the retarded (r) and advanced (a) Green's functions at stroboscopic times can be obtained in the frequency domain as
\begin{equation}\label{G_def}
        \mathcal{G}_{\rm F}^{r(a)}(\omega)= (\omega\pm i\delta - H_{\rm F})^{-1}\,,
\end{equation}
where $\delta$ is an infinitesimally small positive number that helps define the retarded (advanced) nature of the Green's function and $H_{\rm F}$ is the effective Floquet Hamiltonian given by Eq.\,(\ref{Heff}) and written in Nambu space. Due to the basis of the effective Hamiltonian,  $\mathcal{G}_{\rm F}^{r(a)}$ has the following Nambu structure
\begin{equation}
\label{GNambu}
\mathcal{G}_{\rm F}^{r(a)}(\omega) = \begin{pmatrix}
                    G^{r(a)}(\omega) & F^{r(a)}(\omega)
                    \\
                    \bar{F}^{r(a)}(\omega) & \bar{G}^{r(a)}(\omega)
            \end{pmatrix}\,,
\end{equation}
where $G^{r(a)}$ and $F^{r(a)}$ represent the normal and anomalous components of the system Green's function. Here,  $F^{r(a)}(\omega)$  enables the calculation of the superconducting pair amplitudes at stroboscopic times, which is the focus of our discussion below. 

Before going further, we discuss  the allowed superconducting symmetries in our system.   Fermi statistics forces the pair amplitude to be an odd function under the exchange of all quantum numbers of the two electrons forming a Cooper pair. Conventionally, the majority of the literature assumes that Cooper pairs form between electrons at the same time, thus, Cooper pairs turn out to be even in the time coordinates or, equivalently, even in frequency. However, suppose we relax this evenness in time. In that case, Fermi statistics allows the pair amplitude to be either even or odd in time (or frequency) provided that the pair amplitude is antisymmetric under the total exchange of quantum numbers \cite{Berezinskii,Balatsky1992,Baltsky1995,Coleman1997,tanaka2011symmetry,schrieffer1994odd,cayao2019odd,triola2020role}. 
Odd-frequency pairing implies that electrons at different times can form Cooper pairs,  leading to exotic transport phenomena including long-range proximity effect in ferromagnet junctions \cite{Efetov2,Bergeret2001}, anomalous proximity effect \cite{PhysRevB.70.012507,TanakaGolubov2007PRL,Asano2011,chiu2021observation,chiu2023tuning,CHIU2024348}, and paramagnetic responses \cite{Tanaka2005,Higashitani1997,Yokoyama2011,Asano2011}. 
To inspect the pair symmetries, it is important to identify the quantum numbers appearing in the pair amplitudes $F^{r(a)}(\omega)$ and analyze the antisymmetry condition under the exchange of all the quantum numbers, see e. g., Refs.\,\cite{TanakaGolubov2007PRL,
TanakaTanuma2007PRB,odd3,tanaka2011symmetry,cayao2019odd,triola2020role,Tanaka2021,TanakaCayaotheory}.
In this regard, we identify the frequency $\omega$, spin $\sigma$, and spatial coordinates, as the quantum numbers appearing in the pair amplitudes. Thus, taking into account these quantum numbers, we can classify the pairing amplitude into four pair symmetry classes: even-frequency spin-singlet even-parity (ESE), even-frequency spin-triplet odd-parity (ETO), odd-frequency spin-triplet even-parity (OTE), and odd-frequency spin-singlet odd-parity (OSO); see also Table \ref{tab:my_label}. Using retarded Green's functions, the oddness/evenness in frequency $\omega$ is inspected by changing into the advanced Green's function when changing $\omega\leftrightarrow-\omega$ \cite{RevModPhys.91.045005}. It is worth noting that the oddness/evenness in frequency can be studied within Matsubara representation, where $\omega_{n}\leftrightarrow-\omega_{n}$ with  $\omega_{n}$ characterizing Matsubara frequencies \cite{RevModPhys.91.045005}. 

\begin{table}
    \centering
    \begin{tabular}{|c|c|c|c|c|} \hline 
         class&  $\omega\to-\omega$&  $\sigma\leftrightarrow\sigma'$& $i\leftrightarrow j$ &total exchange\\ \hline 
         ESE&  even&  singlet& even &odd\\ \hline 
         ETO&  even&  triplet& odd &odd\\ \hline 
         OTE&  odd&  triplet& even &odd\\ \hline 
         OSO&  odd&  singlet& odd &odd\\ \hline
    \end{tabular}
    \caption{Symmetries of Cooper pair amplitudes allowed by Fermi statistics for an arbitrary superconductor with only frequency, spatial, and spin degrees of freedom. }
    \label{tab:my_label}
\end{table}

Now, we inspect the symmetry of the emerging Cooper pair amplitudes in our model. Because the static superconductor is spin-polarized and the drive does not modify the spin, the resulting spin symmetry is the same as that of the static superconductor, namely, spin-polarized which can be seen as an equal spin-triplet (T). Moreover, $F^{r(a)}$ is a matrix in lattice space, and its elements here we denote by $F_{ij}^{r(a)}$ where we dropped the {\rm F} subscript for simplicity in the notation. Thus, in general, there are two spatial coordinates which correspond to the lattice sites $i,j$, which run from $1$ to   $N$, with $N$ being the last lattice site. Hence, the pair amplitudes inside the system $F_{ij}^{r}$ can be decomposed into even and odd functions under the exchange of $i,j$. In what follows, we focus on the local pair amplitudes $i=j$, which are even in space or parity, have  $s$-wave symmetry, and  are  robust against disorder \cite{tanaka2011symmetry,TanakaGolubov2007PRL,PhysRevB.87.104513,Takagi18}. Given that the superconducting pair amplitude of our interest is expected to exhibit spin-triplet and $s$-wave symmetries, the only possibility for the frequency dependence is to be an odd (O) function of frequency. In the time domain, the odd-frequency dependence is revealed by an odd function under the exchange in time coordinates\,  \cite{tanaka2011symmetry,cayao2019odd,triola2020role,Tanaka2021}. Thus, following simple arguments, we showed that our driven system should develop superconducting pair correlations with an odd-frequency (O), spin-triplet (T), even-parity (E) symmetry, which is sometimes referred to as the OTE symmetry, see e.g., Refs.\,\cite{tanaka2011symmetry,cayao2019odd}. Since our system is already spin-polarized and focuses on local pair amplitudes,   the onsite components of the anomalous Green's function given by Eq.\,(\ref{GNambu}), denoted here by $F_{ii}^{r}$, already have   OTE symmetry. As such, there is no need to decompose the symmetries associated with frequency, spin, and space.  For pedagogical purposes, we denote the OTE pair amplitudes as
\begin{equation}
\label{Fff}
    F_{ii}^{\rm odd}(\omega) =  F_{ii}^r(\omega)\,,
\end{equation}
where $F_{ii}^{r}$ represents  the onsite components of the anomalous Green's function given by Eq.\,(\ref{GNambu}). In what follows we discuss the emergence of the OTE pair amplitudes given by Eq.\,(\ref{Fff})  at stroboscopic times of a time-periodic $p$-wave  superconductor described by the effective Floquet Hamiltonian $H_{\rm F}$ in Eq.\,(\ref{Heff}).

It is worth pointing out that, in  static superconductors without time-periodic drives, the OTE pair amplitude accumulated at the edge is known to be correlated to the topology of the system 
\cite{tanaka2011symmetry,cayao2019odd,triola2020role,Tanaka2021}. In particular,  in the static regime, it has been shown that the odd-frequency pairing acquires a  divergent profile in the presence of MZMs, thus exhibiting a maximum value at the edges of the system and can penetrate deep into it  with a penetration depth similar to that of MZMs. This behaviour implies a strong correlation between  topology and odd-frequency pairing and here we numerically investigate whether this effect persists in the time-driven regime.

\subsection{Accumulated odd-frequency pairing in driven topological superconductors}
To facilitate a meaningful comparison between odd-frequency pairing and Majorana edge modes, we perform a summation of the values of odd-frequency pairing over the leftmost sites \cite{tamura18}. The summation is terminated at the midpoint of the lattice and the summed odd-frequency pairing 
 is   referred to as   ``accumulated odd-frequency pairing at the edge", which is  obtained  as
\begin{equation}
\label{Foddsum}
F_{\rm odd}(\omega)=\sum_{i=1}^{N/2}F_{ii}^{\rm odd}(\omega)\,,
\end{equation}
where $F_{ii}^{\rm odd}$ are diagonal elements in lattice space defined  in Eq.\,(\ref{Fff}). We also fix the initial time such that $t_0 = T/4$ because it allows us to define $F_{\rm odd}$  as an invariant quantity and to establish a straightforward relationship to the topological phases discussed in Section \ref{section3}.

   \begin{figure}[!t]
\centering
	\includegraphics[width=\columnwidth]{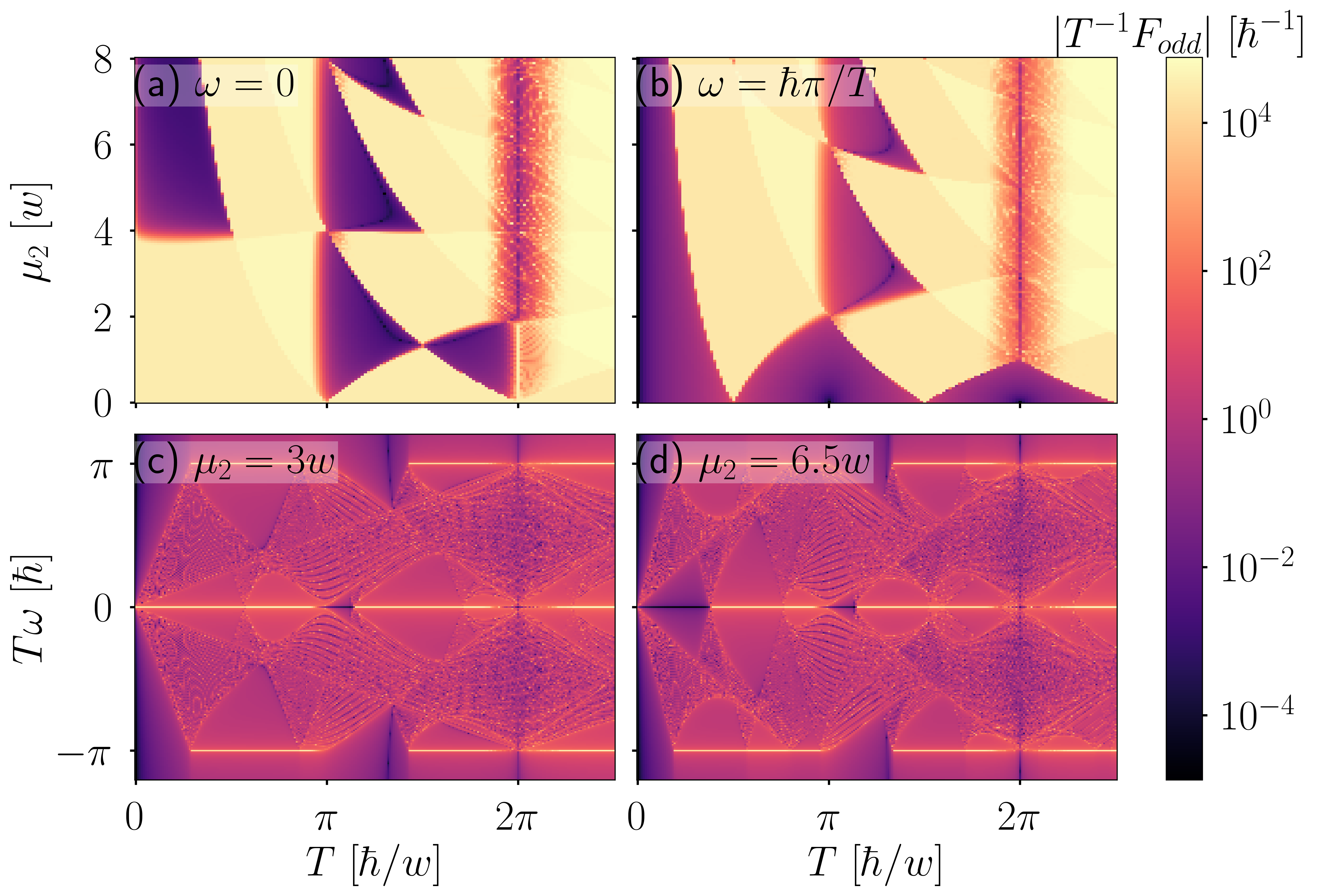}\\
 	\caption{(a,b) Absolute value of the accumulated odd-frequency pair amplitude $F_{\rm odd}$ as a function of $\mu_{2}$ and $T$ for $\omega=0$ and $\omega=\pi\hbar/T$. (c,d) Same as in (a,b) but now as a function of $\omega$ and $T$ for $\mu_2 = 3w$ and  $\mu_{2}=6.5w$.  Parameters: $\Delta=w,\mu_1=0$, $N=100$, $\delta/w=10^{-5}$.}
\label{Fig4} 
\end{figure}

\begin{figure}[!t]
\centering
	\includegraphics[width=\columnwidth]{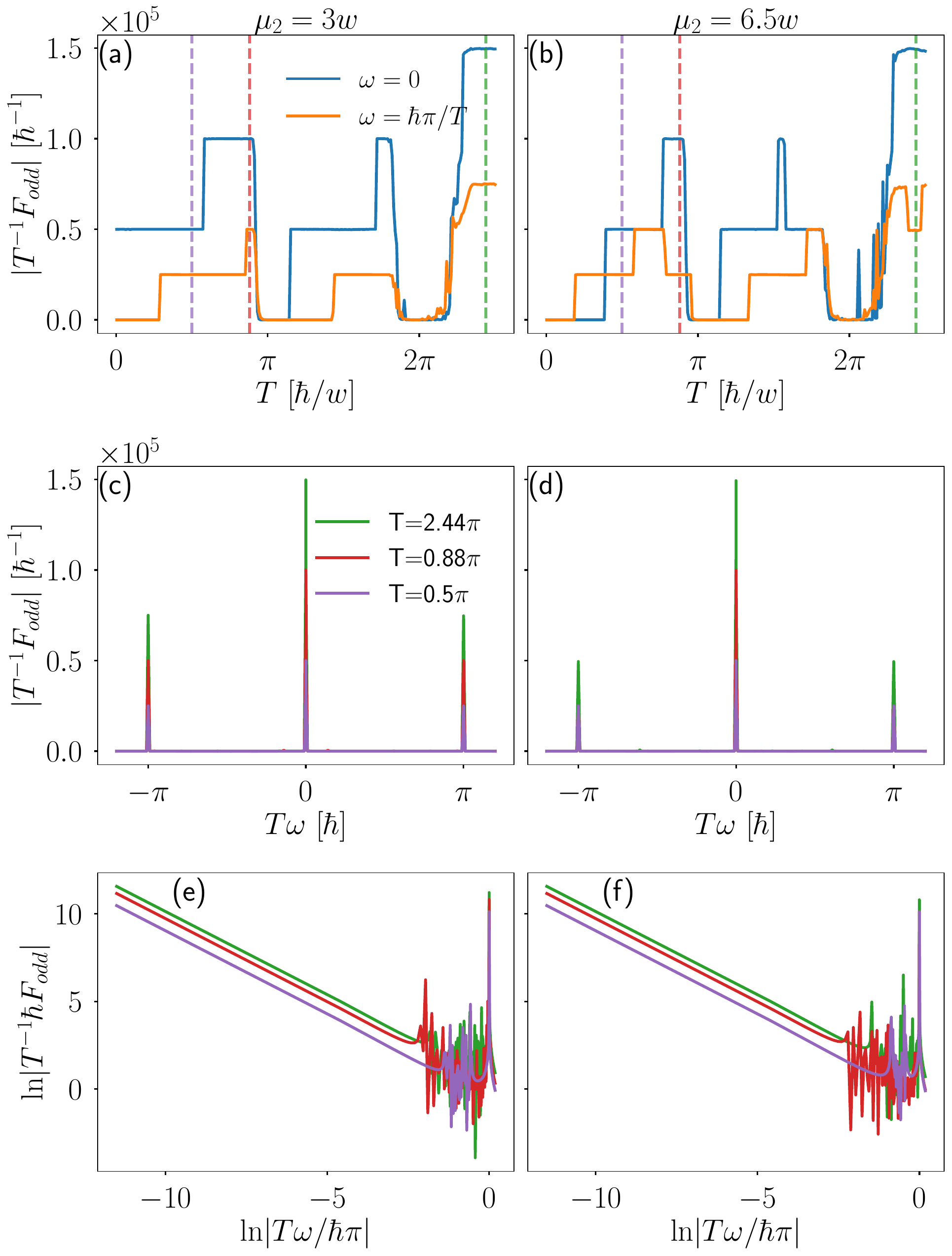}\\
 	\caption{Absolute value of the accumulated odd-frequency pair amplitude $F_{\rm odd}$ as a function of $T$ (a,b), and   $\omega$ (c,d,e,f). The left and right columns correspond to $\mu_{2}=3w$ and $\mu_{2}=6.5w$, respectively.  Parameters: $\Delta=w$, $\mu_{1}=0$, $N=100$, $\delta/w = 10^{-5}$.}
\label{Fig5} 
\end{figure}

In Fig.\,\ref{Fig4}(a,b) we present the magnitude of the  accumulated OTE pair amplitude $|F_{\rm odd}|$ as a function of the chemical potential $\mu_{2}$ and  period $T$ at fixed frequencies $\omega=0$ and $\omega=\pi\hbar/T$ that capture the MZMs and MPMs, respectively. In Fig.\,\ref{Fig4}(c,d) we show the same as in (a,b) but as a function of frequency $\omega$ and $T$ for two fixed chemical potentials $\mu_{2}=3w$ and $\mu_2=6.5w$. The immediate observation is that $|F_{\rm odd}|$ in Fig.\,\ref{Fig4}(a,b)  develops large (vanishing small) values in the range of parameters where the topological invariants shown in Fig.\,\ref{Fig2} are nonzero (zero), suggesting an intriguing relationship between topology and the multiple MZMs and MPMs. This behaviour can be further seen by taking line cuts of Fig.\,\ref{Fig4}(a,b) at fixed chemical potentials $\mu_{2}=3w$ and $\mu_{2}=6.5w$, which is shown in Fig.\,\ref{Fig5}(a,c) and Fig.\,\ref{Fig5}(b,d), respectively.
There we observe  that $|F_{\rm odd}|$ has indeed a similar behavior as the topological invariants in Fig.\,\ref{Fig2}(a,b), being zero where $W_{0,\pi}=0$ and developing plateaus of distinct heights, whose values depend on the number of MZMs or MPMs.  

We note that the resemblance at higher values of $T$ is partially destroyed because of finite-size effect. Due to the small size of the system under study, MZMs and MPMs at the two ends of the chain can hybridize, acquiring nonzero (or non-$\pi$) hybridization energy. This lifts the topological degeneracy, which spoils the correspondence between Majorana states and odd-$\omega$ pairing. Moreover, the localization length is inversely proportional to the quasienergy gap. When the gap is very small, MZMs and MPMs can have localization lengths comparable to the system size. Since we terminate the summation in Eq.\eqref{Foddsum} in the middle of the system, $F_{\rm odd}$ cannot fully capture the effect of MZMs and MPMs if  their localization length is appreciably large relative to the system size. These two finite size effects -hybridization and large localization lengths relative to the system size- are commonly observed for large values of $T$, see for example Fig.\ref{Fig4} near $T=2\pi\hbar/w$. This is because for  large values of  $T$, longer range hoppings and pairing become larger and the localization length becomes longer. Hence, one has to take a larger value of $N$ to obtain a converged value.   We have verified that for a large enough lattice, a one-to-one correspondence exists between topological invariants and $|F_{\rm odd}|$, see next subsection. Therefore, the OTE pair amplitude enables a full mapping of the topological invariants in the presence of the time-periodic drive.
At this point we point out that a  relationship between MZMs and topological invariants has been  discussed before but only in the static regime \cite{tamura18}. Our results discussed in this part   thus demonstrate that  such a relationship persists even in the driven phase involving multiple MZMs and MPMs.

\begin{figure}[!t]
    \centering
    \includegraphics[width=1\columnwidth]{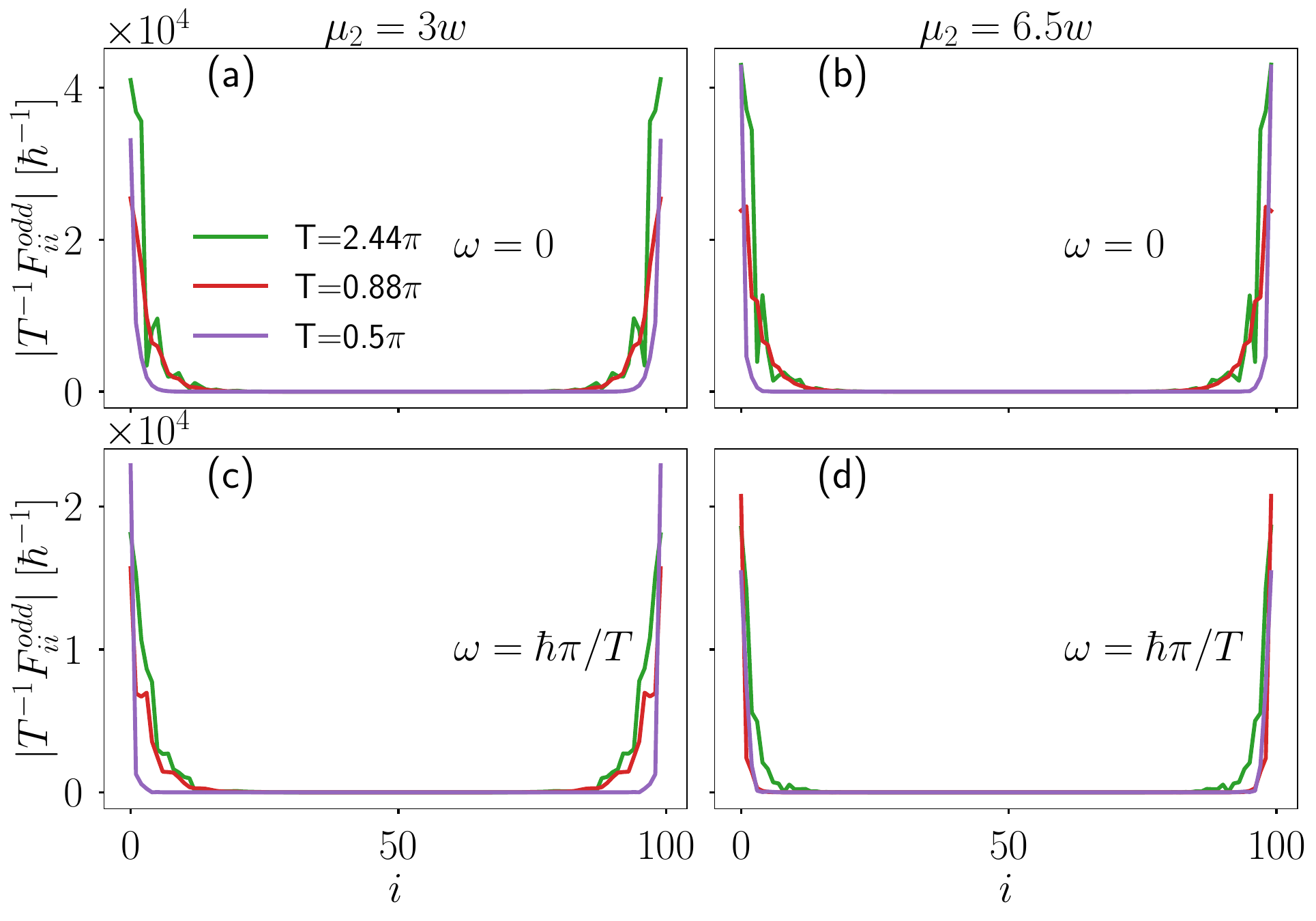}
    \caption{Onsite odd-frequency pair amplitude $F_{ii}^{\rm odd}$  as a function of space $i$. The left and right columns correspond to $\mu_{2}=3w$ and $\mu_{2}=6.5w$, respectively. Parameters: $\Delta=w$, $\mu_{1}=0$, $N=100$, $\delta/w = 10^{-5}$.}
    \label{Fig5-2}
\end{figure}
Further insights on the relationship between Majorana modes and OTE pair correlations can be seen in Fig.\,\ref{Fig4}(c,d) where we present $|F_{\rm odd}|$ as a function of $\omega$ and $T$. Interestingly, the odd-frequency pair amplitude  exhibits its largest values at $\omega=0$ and $\omega=\pi\hbar/T$, which correspond to the quasienergies of MZMs and MPMs, respectively, discussed in Section \ref{section3b} and Fig.\,\ref{Fig3}.  
This implies that  the odd-frequency pairing resonates at the quasienergies of the MZMs and MPMs,   acquiring divergent frequency profiles that follow $\sim1/\omega$  and $\sim1/(\omega-\pi\hbar/T)$. This intriguing behaviour is confirmed in Fig.\,\ref{Fig5}(c,d), where the odd-frequency pair amplitude $|F_{\rm odd}|$ is plotted as a function of $\omega$ at three fixed values of $T$. We also show ${\rm log}|T^{-1}\hbar F_{\rm odd}|$ in  Fig.\,\ref{Fig5}(e,f), which clearly reveals the  divergent profile of $|F_{\rm odd}|$ at $\omega=0$ and $\omega=\pi\hbar/T$: here, the divergent peak at ${\rm log}|T\omega/\hbar\pi|=0$ depicts the divergence of the OTE pair amplitude  at $\omega=\pi\hbar/T$, while the linear behavior shows that the OTE pair amplitude is also divergent when ${\ln}|T\omega/\hbar\pi|$ takes large negative values (i.e. as $\omega$ approaches zero) \footnote{For ${\ln}|T\omega/\hbar\pi|=-12$, which is roughly the leftmost value of the $x$ axis in Fig.\,\ref{Fig5}(e,f), we obtain $\omega\approx10^{-6}(\hbar \pi/T)$ and $F_{\rm odd}\approx 10^{5}(T/\hbar)$, thus revealing the enhancement of the OTE amplitude at zero quasienergy.}.  The multiple peak structure appearing  for ${\ln}|T\omega/\hbar\pi|<0$ corresponds to contributions coming from the quasicontinuum of states seen in the dense regions of Fig.\,\ref{Fig3}(a,b);   they are, however, much smaller than OTE pairing due to MZMs and MPMs, as seen in Fig\,\ref{Fig5}(c,d). Furthermore,   we note that the OTE amplitude develops larger values when the number of MZMs or MPMs is larger, as seen by noting the green, red, and magenta colors in e.g., Fig.\,\ref{Fig5}(c,d).   

Having explored the frequency dependence of the OTE pair amplitude, we now inspect its real space dependence to gain an even better understanding of its behaviour under the presence of Majorana modes. For this purpose, we address the odd-frequency pair amplitude before performing the summation in Eq.\,(\ref{Foddsum}), namely, we inspect  the real space profile of $F_{ii}^{\rm odd}$. This is visualized in Fig.\,\ref{Fig5-2}, where we plot the absolute value of the OTE pair amplitude $|F_{ii}^{\rm odd}|$ as a function of $i$ at $\omega=0$ (a,b) and $\omega=\pi\hbar/T$ (c,d); here the top and bottom rows correspond to $\mu_{2}=3w$ and $\mu_{2}=6.5w$, respectively. The main feature of the space dependence of the OTE pair amplitude is that they exhibit maxima at the edges of the system, in a very similar way as expected from the spatial nonlocal wavefunctions of Majorana edge modes. The accumulation of large OTE pair correlations at the edges is a well-established effect in the static regime of topological superconductors, which reinforces the strong connection between MZMs and OTE pairing \cite{tanaka2011symmetry,cayao2019odd,triola2020role,Tanaka2021,TanakaCayaotheory}. Our findings therefore reveal that the relationship between odd-frequency pairing and topology also holds for driven superconductors containing multiple MZMs and MPMs. Consequently, the nature of the emergent superconducting pairs at the boundaries of time-periodic topological superconductors is of OTE symmetry.

\subsection{Characterization of topology by the odd-frequency pairing amplitude}
\label{section4b}
\begin{figure*}[!th]
    \centering
    \includegraphics[width=\linewidth]{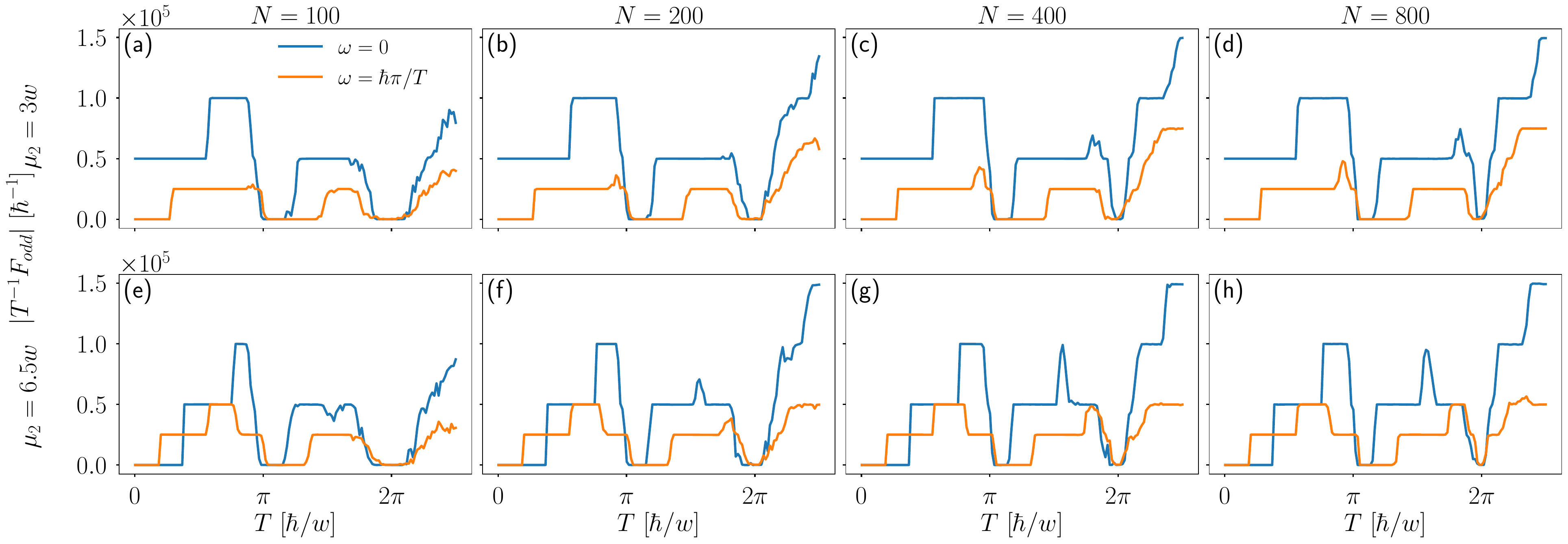}
    \caption{Same as in Fig. \ref{Fig5}(a,b) but in the presence of on site disorder and for chain length of 100 sites(a,e), 200 sites(b,f), 400 sites(c,g), 800 sites(d,h). Parameters: $V_0=w, \Delta=w, \mu_1=0$, $\delta/w = 10^{-5}$}
    \label{fig:disorder}
\end{figure*}
So far we have seen that there is an intriguing relationship between the emergence of Majorana edge states and odd-frequency  pairing with OTE symmetry in time-periodic topological superconductors.  Here we further investigate that this relationship is not accidental but it   stems from the so-called spectral bulk boundary correspondence (SBBC), which predicts that the bulk winding number is related to  odd-frequency pairing at boundaries in chiral symmetric systems \cite{tamura18,PhysRevB.100.174512,PhysRevB.101.214507}.  The SBBC was studied before in chiral symmetric static systems and  here we explore it in the time-periodic domain.

We inspect the accumulated odd-frequency pairing $F_{\rm odd}$ defined in Eq.\,(\ref{Foddsum}) with  $t_{0}= T/4$. For this purpose, we denote $     F_{\rm odd}^{\rm N} (z)\equiv   F_{\rm odd}(z)$ and write
\begin{equation}\label{eq:F_trace}
     F_{\rm odd}^{\rm N} (z)=\sum_{i=1}^{N/2}F_{ii}^{\rm odd}=\frac{1}{4}{\rm \hat{T}r}\{\Gamma \mathcal{G}(z)+ \mathcal{G}(z)\Gamma\}\,,
\end{equation}
where $\Gamma=\tau_{x}$ is the chiral operator, $\mathcal{G}(z)=(z-H_{\rm F})^{-1}$ is the Green's function,   ${\rm \hat{T}r}$  represents a special trace operation that sums up the contributions from the leftmost lattice sites up to the middle of the system only, and $z$ represents complex frequencies, whose analytic continuation $z\rightarrow \omega\pm i\delta$ gives the advanced  and retarded Green's functions studied in Eq.\,(\ref{Foddsum}).

 The right hand side of Eq.\,\eqref{eq:F_trace} makes chiral symmetry of the system manifest and highlights the connection between odd-$\omega$ pairing and chiral symmetry. More interestingly, we can use Eq.\,\eqref{eq:F_trace} to study the universal behaviour of odd-$\omega$ pairing in chiral symmetric Floquet systems. In particular, we have analytically confirmed that for any chiral Floquet Hamiltonian, odd-$\omega$ pairing follows a universal profile in the presence of Majorana states. This can be understood by taking the system size $N$ very large  such that the system becomes semi-infinite, for the limit  $z\rightarrow0$ or $z\rightarrow \pi\hbar/T$, we find 
\begin{equation}
\label{SBBC}
\lim_{N\rightarrow\infty} F_{\rm odd}^{\rm N} (z)\xrightarrow{z\rightarrow E_\omega}\frac{1}{2}\frac{ W_{\omega}}{z-E_{\omega}}\,,
\end{equation}
where $W_{\omega}$ corresponds to the winding numbers $W_{0(\pi)}$ obtained from Eq.\,(\ref{W0pi}), respectively. Moreover, $E_{\omega}$ corresponds to quasienergies $E_{0}=0$ and $E_{\pi}=\pi\hbar/T$ of MZMs and MPMs, respectively.  For details on the derivation, see Appendix \ref{app:c} and Appendix \ref{app:e}. The accumulated odd-frequency pairing in Eq.\,(\ref{SBBC}) diverges at quasienergies of MZMs or MPMs and develops distinct heights proportional to the winding numbers, in agreement with what we found in Fig.\,\ref{Fig5}. 
 We note that in Fig.\,\ref{Fig5}, the size of the peak for $\omega=\pi\hbar/T$ is half of what it is expected to be at according to Eq.\,\eqref{SBBC}. This is because our numerical simulation treats $\omega=\pi\hbar/T$ and $\omega=-\pi\hbar/T$ as two different frequencies, while in reality, they are the same value due to periodicity. Thus, our numerical program can only count half of the actual value. Nevertheless, Eq.\,(\ref{SBBC}) clearly demonstrates  the direct relationship between   odd-frequency  pairing and topology in time-periodic topological superconductors.  Also, it is worth noting that Eq.\,(\ref{SBBC}) extends the SBBC at low frequencies, initially investigated for static systems \cite{tamura18,PhysRevB.100.174512,PhysRevB.101.214507}, to the time-dependent domain.

We emphasize that, to derive Eq.\,(\ref{SBBC}), the only assumption is  that the one-period evolution operator is chiral symmetric, disregarding additional details of the Hamiltonian. Hence, the accumulated odd-frequency pairing and winding number in Eq.\,(\ref{SBBC}) is protected by chiral symmetry and thus, it is stable under perturbations that preserve the symmetry. Therefore, we expect that Eq.\eqref{SBBC} is a generic relationship applicable to all one-dimensional chiral or sublattice symmetric systems (AIII, BDI, CII,  and DIII in the 10-fold classification \cite{RevModPhys.88.035005}), including Rashba systems \cite{TanakaCayaotheory} and helical Shiba chains \cite{hsu2021helical}. Similarly, we expect that Eq.\,(\ref{SBBC}) is applicable to other  time-periodic drives as long as chiral symmetry is protected. We can therefore regard the   accumulated odd-$\omega$ pairing discussed here as a real-space representation of the winding number, which is    equivalent to   other real space representations of the winding number \cite{PhysRevB.103.224208,PhysRevLett.113.046802}; see also Ref.\,\cite{PhysRevB.100.174512} for a proof in the static case.

\subsection{Robustness of the odd-frequency pairing against scalar disorder}
To demonstrate the stability of Eq.\eqref{SBBC} under realistic conditions, we here compute the emergent odd-frequency pairing under chiral symmetry-preserving scalar disorder. For this purpose, we modify the time-periodic Hamiltonian $H(t)$ by adding a random onsite potential,
\begin{equation}
    H_{\rm dis}(t) = H(t) + \sum_{i=1}^N\psi_i^\dagger(v_i\tau_z)\psi_i\,,
    \label{eq:H_disorder}
\end{equation}
where $v_1,\dots, v_N$ are random onsite energies uniformly distributed inside the region $[-\frac{V_0}{2},\frac{V_0}{2}]$ with  $V_0$ being the  strength of disorder. Next, we investigate the accumulated odd-$\omega$ pairing amplitude $F_{\rm odd}$ for the disordered Hamiltonian $H_{\rm dis}$. We focus on the case of intermediate disorder by setting $V_0 = w$ and present $F_{\rm odd}$ averaged over 10 realizations of disorder; we have verified that more disorder realizations do not affect the shown results.

In Fig. \ref{fig:disorder}  we show the disorder averaged $F_{\rm odd}$ as a function of the period $T$ for different lengths of our system  and for the two values of $\mu_{2}$ discussed in Fig. \ref{Fig5}. Note that Fig. \ref{fig:disorder}(a,e) corresponds to a disordered system of 100 lattice sites, which is the   same length as in  Fig. \ref{Fig5} but therein without   disorder. 
The first feature to notice is that, for smaller periods and small systems, the correspondence between odd-$\omega$ pairing amplitude and the winding numbers is clearly visible even in the presence of intermediate disorder, see Fig. \ref{fig:disorder}(a,e) and Fig. \ref{Fig3}(c,d). However, the one-to-one correspondence between odd-$\omega$ pairing and the winding numbers is weakened for larger $T$ due to the finite size effect in small systems. In fact, as the length of the system increases, the correspondence improves and we can clearly identify the different topological phases and the topological transition points purely from $F_{\rm odd}$, see  Fig. \ref{fig:disorder}(d,h).

Generally speaking, we expect Eq.\eqref{SBBC} to hold as long as an quasienergy gap separating the Majorana states from the bulk exists. For strong  disorder, we expect that the bulk quasienergy modes will not be separated from the Majorana modes by an quasienergy gap and thus the winding number becomes ill-defined and Eq. \eqref{SBBC} does not hold. Nevertheless, our results indicate that Eq.\,\eqref{SBBC} is topologically-protected and is robust against scalar disorder which protects chiral symmetry. We expect our relationship to hold for all topological classes with chiral symmetry  (AIII, BDI, CII, and DIII in the 10-fold classification \cite{RevModPhys.88.035005}) in one dimension.

\section{Concluding remarks}
\label{section5}
In conclusion, we considered a one-dimensional $p$-wave superconductor under a time-periodic chemical potential and studied the relationship between the emergence of Majorana edge modes and odd-frequency pairing. We found that  Majorana modes appear   located not only at quasienergy $E_{\rm F}=0$, but also at $E_{\rm F}=\pi\hbar/T$, and the number of such Majorana edge modes can be controlled by the driving period and chemical potentials, in agreement with previous studies. Interestingly, we discovered that odd-frequency pairing is finite and acquires large values in the presence  of Majorana  edge zero and $\pi$ modes, representing a unique out-of-equilibrium phenomenon that can be fully controlled by the drive. In particular, we found that the frequency dependence of the formed odd-frequency pairing in the presence of Majorana zero and $\pi$ edge modes  is proportional to $\sim1/\omega$ and $\sim1/(\omega-\pi\hbar/T)$, respectively. Moreover,  we showed that the divergent profile of odd-frequency pairing remains even in the presence of multiple Majorana edge modes  and scalar disorder. We have also established a bulk-boundary correspondence between boundary odd-$\omega$ pairing and bulk topology, which only requires chiral symmetry and therefore applicable to other chiral symmetric Floquet superconductors under distinct types of chiral symmetric drives. This bulk-boundary correspondence reveals that the boundary odd-frequency pairing can be regarded as a real space  topological invariant, providing a powerful tool to characterize topology in time-periodic superconductors even in the presence of disorder.  Our results thus establish an intimate relationship between Majorana modes and odd-frequency pairing  in time-driven topological superconductors.  

Furthermore, our results offer a dynamical approach to control odd-frequency pairing, which could be relevant for superconducting spintronics where spin-polarized Cooper pairs are used \cite{linder2015superconducting}. Similarly, the feasibility of realizing the driven odd-frequency pairing we predict relies on the  physical implementations of the Kitaev chain. In this regard, there have been enormous advances in superconductor-semiconductor hybrid systems where topological superconductivity is expected at large magnetic fields and controlled by the system parameters \cite{TanakaCayaotheory}. Although there still are unavoidable challenges \cite{PhysRevB.91.024514,PhysRevB.104.L020501,prada2019andreev,PhysRevB.104.134507}, recent advances \cite{lutchyn2018majorana,prada2019andreev} suggest that   topological superconductivity is very likely to be achieved soon. Our findings   therefore hold experimental relevance.

It is worth noting that our findings also open further questions in time-periodic topological superconductors. For instance,  while our results are valid for all chiral symmetric Hamiltonians, it is interesting to study the relationship between topology and odd-frequency pairing in more realistic setups such as Rashba nanowires \cite{TanakaCayaotheory} or helical Shiba chains \cite{hsu2021helical}.  
Another interesting aspect is the relationship between odd-frequency pairing and topology in dissipative time-periodic topological superconductors, specially in relation to the non-Hermitian skin effect since it appears at the boundary \cite{PhysRevX.9.041015}. Similarly, it would be interesting to explore the validity of the bulk boundary correspondence we demonstrate here in the presence of many-body interactions as well as under the effect of multiple frequencies, which could be relevant for establishing odd-frequency time-crystalline superconducting phases.


\section{Acknowledgements} 
 We thank  S. Ikegaya and Y. Kawaguchi  for insightful discussions.   
E. A. acknowledges financial support from Nagoya University and Mitsubishi Foundation. S. T. thanks the support of   the W{\"u}rzburg-Dresden Cluster of Excellence ct.qmat, EXC2147, project-id 390858490, the DFG (SFB 1170), and the Bavarian Ministry of Economic Affairs, Regional Development and Energy within the High-Tech Agenda Project ``Bausteine f{\"u}r das Quanten Computing auf Basis topologischer Materialen''.  Y. T. acknowledges support from JSPS with Grants-in-Aid for Scientific research  (KAKENHI Grants No. 23K17668  and 24K00583). J. C. acknowledges financial support from the Japan Society for the Promotion of Science via the International Research Fellow Program,  the Swedish Research Council (Vetenskapsr{\aa}det Grant No. 2021-04121),  and the Carl Trygger’s Foundation (Grant No. 22: 2093).

 
\appendix
\renewcommand{\thepage}{A\arabic{page}}
\setcounter{page}{1}
 \renewcommand{\thefigure}{A\arabic{figure}}
\setcounter{figure}{0}

\section{Bulk quasienergy and bulk gaps}
\label{AppendixB}
In this part, we derive the bulk quasienergy given by Eq.~\eqref{quasienergy}. For convenience, without loss of generality, we set $t_0=0$ in Eq.\,(\ref{UTt0}) and write the one-period time-evolution operator in momentum space in the following form
\begin{equation}\label{eq:ap_B-UT}
    U_T = e^{-i\frac{T}{2\hbar}\bm{E}_2\cdot\bm{\tau}}e^{-i\frac{T}{2\hbar}\bm{E}_1\cdot\bm{\tau}}\,,
\end{equation}
with $\bm{E}_j= (0,-2\Delta\sin{k},-\mu_j-2w\cos{k})^T$ and $\bm{\tau}=(\tau_{x},\tau_{y},\tau_{z})$ the vector of Pauli matrices in Nambu space.  As we see, the one-period time-evolution operator is expressed as a product of exponentials of Pauli matrices, which implies that it  should follow the group composition law of $SU(2)$, i.e., it can be written as an exponential of $\mathfrak{su}(2)$ algebra. Thus, using Euler's formula for Pauli matrices, we can write Eq.\,(\ref{eq:ap_B-UT}) as
\begin{widetext}
\begin{equation}
\label{UTmatrixdecomp}
\begin{split}
    U_T &= \Bigg[\cos\frac{E_1T}{2\hbar}\cos\frac{E_2T}{2\hbar}
    - \hat{E}_1\cdot\hat{E}_2\sin\frac{E_1T}{2\hbar}\sin\frac{E_2T}{2\hbar}\Bigg]  \\
&    +i\Bigg[\hat{E}_1\sin\frac{E_1T}{2\hbar}\cos\frac{E_2T}{2\hbar}
 +\hat{E}_2\sin\frac{E_2T}{2\hbar}\cos\frac{E_1T}{2\hbar} 
  +\hat{E}_1\times\hat{E}_2\ \sin\frac{E_1T}{2\hbar}\sin\frac{E_2T}{2\hbar}\Bigg]\cdot\bm{\tau}\,,
\end{split}
\end{equation}
\end{widetext}
where $\hat{E}_j= {\bm{E}_j}/{E_j}, E_j=|{\bm E}_j|$, and ${\bm E}_j$ is defined below Eq.\,(\ref{eq:ap_B-UT}).  Then, we use the fact that bulk quasienergies are eigenvalues of the eigenvalue problem defined by
$U_{\rm T}\ket{\psi}=e^{-i\frac{T}{\hbar}H_{\rm F}(k)}\ket{\psi}$, where $H_{\rm F}(k)=\bm{E}_{\rm F}\cdot\bm{\tau}$. Thus, 
\begin{equation}
\label{HFmatrixdecomp}
e^{-i\frac{T}{\hbar}H_{\rm F}(k)}=\cos\frac{E_{\rm F}T}{\hbar} + i\hat{E}_{\rm F}\cdot\bm{\tau}\sin\frac{E_{\rm F}T}{\hbar} \,,
 \end{equation}
 At this point, we equate the Eq.\,(\ref{UTmatrixdecomp}) and Eq.\,(\ref{HFmatrixdecomp}), and, by comparing their terms proportional to the identity, we find that the Floquet quasienergy is given by 
 \begin{equation}
 \label{EFloquet}
\begin{split}
    E_{\rm F}(k) &= \frac{\hbar}{T}\arccos \bigg[\cos(E_1T/2\hbar)\cos(E_2T/2\hbar) \\ 
    &- \frac{\bm{E}_1\cdot\bm{E}_2}{E_1E_2}\sin(E_1T/2\hbar)\sin(E_2T/2\hbar) \bigg]\,.
    \end{split}
\end{equation}
This is the Floquet bulk quasienergy presented in Eq.~\eqref{quasienergy} of Subsection \,\ref{section3b}.   We note that the expression given by Eq.\,(\ref{EFloquet}) is general as   long as the one-period time-evolution operator takes the form of Eq.\,(\ref{eq:ap_B-UT}), suggesting its usefulness in other time-periodic systems.

One particular point here is that, at $k=\pi$, Eq.\,(\ref{EFloquet}) transforms into ${\rm cos}[TE_{\rm F}/\hbar]={\rm cos}[T(E_{1}+E_{2})/(2\hbar)]={\rm cos}[T(|-\mu_{1}+2w|+|-\mu_{2}+2w|)/(2\hbar)]$. Then, for $\mu_{1}+\mu_{2}=4w$, we get   ${\rm cos}[TE_{\rm F}/\hbar]=1$, which gives quasienergies equal to $E_{\rm F}=(2\pi n)\hbar/T$ for $n\in \mathbb{Z}$. Hence, for $n=0$, the quasienergy vanishes $E_{\rm F}=0$, which can be seen as a gap-closing feature coming from the undriven phase because it happens at $\mu_{1}+\mu_{2}=4w$ which is the topological phase transition in the undriven regime where the system is described by $(H_{1}+H_{2})/2$. This gap closing is discussed in Subsection \ref{section3a}.

\section{Baker-Campbell-Hausdorff  expansion of the Floquet Hamiltonian}
\label{AppendixA}
In this part, we discuss the expansion of the Floquet Hamiltonian $H_{\rm F}$ given by Eq.\,(\ref{Heff}) in terms of the driving period $T$. This is of particular relevance to understand  the topological phase transition at $T=0$ and also the emergence of multiple Majorana modes due to long-range hopping, both in Sec\,\ref{section3}. For this purpose,  we fix $t_{0}=0$ without loss of generality  and perform a Baker-Campbell-Hausdorff  expansion of $H_{\rm F}$ up to third order in $T$, obtaining
\begin{equation}
H_{\rm F}(k)=\sum_{j=1}^{\infty} h_j\left(\frac{T}{\hbar}\right)^{j-1},
\end{equation}
where
\begin{widetext}
\begin{equation}
\label{ElementsBCH}
\begin{split}
  h_1 
    &=
    \frac{1}{2}(H_1+H_2)\\
   & =
    -[(\mu_1+\mu_2)/2+2w\cos(k)]\tau_z-2\Delta\sin(k)\tau_y\,,\\
 h_2
    &=
    \frac{1}{2}\frac{1}{2^2}[H_1,H_2]\\
    &=
    \frac{i}{2}\Delta(-\mu_1+\mu_2)\sin(k) \tau_x\,,   \\
    h_3
    &=
    \frac{1}{12}\frac{1}{2^3}([H_1,[H_1,H_2]]+[H_2,[H_2,H_1]])\\
    &=
    \frac{-1}{12}\Delta{(\mu_1-\mu_2)}^2\sin(k)\tau_y\,,\\
    h_4
    &=
    -
    \frac{1}{24}\frac{1}{2^4}[H_2,[H_1,[H_1,H_2]]]\\
    &=
    \frac{i}{24}\Delta(\mu_1-\mu_2)\sin(k)
    \big[-2\Delta^2-\mu_1\mu_2-2w^2-2(\mu_1+\mu_2)w\cos k+2(\Delta^2-w^2)\cos(2k)\big]\tau_x\,,
\end{split}
\end{equation}
\end{widetext}
where $H_{1,2}$ are given by Eq.\,(\ref{eq:Hamiltonian}) while $\Delta$, $w$, and $\mu_{1,2}$ are the order parameter, hopping, and chemical potentials respectively, as discussed in Eqs.\,(\ref{eq1}) and (\ref{eqmut}).

The first observation in Eqs.\,(\ref{ElementsBCH}) is that the lowest term (near $T=0$) is given by $h_{1}=(H_{1}+H_{2})/2$, implying that the phase transition at $T=0$ is given by $|\mu_{1}+\mu_{2}|/2=2w$. Hence, for $\mu_{1}=0$, the topological phase transition is given by $|\mu_{2}|=4w$ as we indeed obtain in   Subsection \ref{section3a} and Fig.\,\ref{Fig2}. The second observation is the emergence of momentum-dependent terms proportional to $[{\rm sin}(2k)]\tau_{x}$ and $[{\rm sin}(k){\rm cos}(2k)]\tau_{x}$ in the fourth-order correction $h_{4}$. Thus, when performing a Fourier transformation to real space, these elements give rise to longer range pairing (next-nearest neighbor and second-next-nearest neighbor), which is used to argue the emergence of multiple Majorana modes in Subsection \ref{section3b}. By adding more orders in the expansion, it is possible to also see the emergence of long-range hopping.

   \section{Bulk Boundary correspondence and odd-$\omega$ pairing}\label{app:c}
   In this appendix, we derive Eq.\,\eqref{SBBC}  for a wide class of one-dimensional Hamiltonians and driving protocols. The only ingredient for the proof is the existence of chiral symmetry in the form of a unitary operator $\Gamma[t_0]$ which depends on the initial time satisfying $\Gamma[t_0]^2= 1$ such that the one-period evolution operator $U_T[t_0]$ satisfies
   \begin{equation}\label{eq:chiral symmetry}
       \Gamma[t_0] U_T[t_0]\Gamma[t_0] = U_T^\dagger[t_0]\,,
   \end{equation}
for some initial time $t_0$. We note that the time-periodic Hamiltonian need not satisfy chiral symmetry as long as Eq.\,\eqref{eq:chiral symmetry} is satisfied. The chiral symmetry can be engineered via driving even in the absence of the symmetry in the time-periodic Hamiltonian (see for example \cite{PhysRevB.87.201109}). 

We define the following function of complex frequency $z$
\begin{equation} \label{eq: F_z appndx}
    \begin{split}
        F&(z) = \frac{1}{4}{\rm \hat{T}r}\left\{{\Gamma[t_0]\mathcal{G}_{\rm F}[t_0](z) + \mathcal{G}_{\rm F}[t_0](z) \Gamma[t_0]}\right\}
    \\
    &= \frac{1}{4}{\rm \hat{T}r}\left\{{\Gamma[t_0]\frac{1}{z-H_{\rm F}[t_0]} + \frac{1}{z-H_{\rm F}[t_0]} \Gamma[t_0]}\right\} \,,
    \end{split}
\end{equation}
where the trace is taken over the leftmost sites on the chain and the summation is terminated in the middle. $F(z)$ is odd in $z$ and can be interpreted as an odd-$\omega$ correlation\cite{tamura18}. In our system, $F(z)$ reduces to Eq.\eqref{Foddsum} when $t_0=\pm\frac{T}{4}$ and $z=\omega+i\delta$; however, $F(z)$ can be defined for any chiral Floquet Hamiltonian. In chiral superconducting systems, $F(z)$ corresponds to the accumulated odd-$\omega$ Cooper pairing amplitude at the left edge but $F(z)$ can also be studied in non-superconducting systems with chiral or sublattice symmetry such as the Su–Schrieffer–Heeger (SSH) model\cite{PhysRevLett.42.1698} where it corresponds to an accumulated odd-$\omega$ charge density wave order parameter(c.f. \cite{PhysRevB.104.165125,PhysRevB.101.214507}.)  

We claim that for a sufficiently large system size, the residues of $F(z)$ at $z=0$ and $z=\frac{\hbar\pi}{T}$ count the difference of the number of the positive chirality and negative chirality zero and  ($\frac{\hbar\pi}{T}$) quasienergy edge modes at the leftmost edge, respectively\cite{Gurarie2011}. According to the index theorem, this difference equals the winding number up to a sign\cite{tanaka2011symmetry,ChiralFloquetWindingNumber}. First, we prove our claim for the residue of Eq.\eqref{eq: F_z appndx} at $z=0$. We express the partial trace in Eq.\eqref{eq: F_z appndx} in terms of the position basis $\ket{j,\sigma}$ where $j$ is the position and $\sigma$ are the internal degrees of freedom. We have
\begin{equation}
    \begin{split}
        \lim_{z\to0}&zF(z)= \lim_{z\to0}\frac{1}{4} \sum_{j=1,\sigma}^{j=N/2}
        \\
        &\expval{\Gamma[t_0]\frac{z}{z-H_{\rm F}[t_0]} + \frac{z}{z-H_{\rm F}[t_0]} \Gamma[t_0]}{j,\sigma}\,,
    \end{split}
\end{equation}
Next, we insert the full resolution of the identity $\sum_E \ketbra{E}$ twice as follows:
\begin{equation}
    \begin{split}
        \lim_{z\to0}&zF(z) = \lim_{z\to0}\frac{1}{4}\sum_{j,\sigma,E,E'}^{j=N/2} \braket{j,\sigma}{E}
        \\
        \times&\mel{E}{\Gamma[t_0]\frac{z}{z-E'} + \frac{z}{z-E} \Gamma[t_0]}{E'}\braket{E'}{j,\sigma}\,.
    \end{split}
\end{equation}
The chiral operator has nonzero matrix elements only when $E=-E'$, thus, most of the terms in the above summation will vanish. Moreover, we now split the summation over the quasienergies into two parts. The first summation is over the  states with $E\approx0$ and the second summation is over everything else. 
Thus, Eq.\eqref{eq: F_z appndx} becomes
\begin{widetext}
    \begin{equation}
        \begin{split}
            \lim_{z\to0}zF(z)&= \lim_{z\to0}\sum_{j,\sigma,E,E'}^{j=N/2}\frac{1}{4} \braket{j,\sigma}{E}\mel{E}{\Gamma[t_0]\frac{z}{z-E'} + \frac{z}{z-E} \Gamma[t_0]}{E'}\braket{E'}{j,\sigma}
            \\
            &=\lim_{z\to0} \sum_{j,\sigma,E\approx0}^{j=N/2}\frac{1}{2}  \braket{j,\sigma}{E}\mel{E}{\frac{z^2}{z^2-E^2} \Gamma[t_0]}{-E}\braket{-E}{j,\sigma}
            \\
            &+  \lim_{z\to0} \sum_{j,\sigma,E\not\approx0}^{j=N/2}\frac{1}{2}  \braket{j,\sigma}{E}\mel{E}{\frac{z^2}{z^2-E^2} \Gamma[t_0]}{-E}\braket{-E}{j,\sigma}\,.
        \end{split}
    \end{equation}    

\end{widetext}
Since the states with \( E \not\approx 0 \) are separated from zero quasienergy by a bulk gap, the second term has a negligible contribution near \( z=0 \). Therefore, the first term dominates for small \( z \).

For large system sizes, the residue of \( F(z) \) at \( z=0 \) is determined by the first term, which sums over the states whose eigenvalues \( E \) are close to zero. As the system size increases, these states approach exactly zero quasienergy, and they localize near the left edge of the system. Thus, for a sufficiently large system, Eq.\eqref{eq: F_z appndx} has nonzero contributions only from the zero-quasienergy states localized on the left edge.

Labeling the zero-quasienergy modes on the left edge as \( \ket{E=0,\text{left edge}} \), we take the limit as the system size grows to infinity:
\begin{widetext}
\begin{equation}
    \begin{split}
        \lim_{z\to0}zF(z) & \xrightarrow{N\to\infty}
        \lim_{z\to0} \sum_{j,\sigma,E=0}^{j=N/2}\frac{1}{2}  \abs{\braket{j,\sigma}{E}}^2\mel{E}{\frac{z^2}{z^2-E^2} \Gamma[t_0]}{-E}
            \\
        &= \sum_{E = 0} \frac{1}{2}\expval{\Gamma[t_0]}{E=0,\text{left edge}}
         \\
        &=\frac{N^{+}_0 - N^{-}_0}{2} =\frac{W_0}{2} \,,
    \end{split}
\end{equation}
\end{widetext}
where \( N_0^{\pm} \) is the number of zero-quasienergy states with positive (negative) chirality at the left edge. In the last line, we used the index theorem to relate the winding number \( W_0 \) to the index \( N^{+}_0 - N^{-}_0 \) \cite{ChiralFloquetWindingNumber}. Thus, we have proven that the residue of \( F(z) \) at \( z=0 \) is proportional to the winding number.

Following similar steps, we can show that the residue of \( F(z) \) at \( z=\frac{\hbar\pi}{T} \) is proportional to the winding number of \( \pi \)-modes as follows:

\begin{widetext}
\begin{equation}
    \begin{split}
        \lim_{z\to\frac{\hbar\pi}{T}}(z-\frac{\hbar\pi}{T})F(z) &\xrightarrow{N\to\infty}        \lim_{z\to\frac{\hbar\pi}{T}} \sum_{j,\sigma,E=\frac{\hbar\pi}{T}}^{j=N/2}\frac{1}{2}  \braket{j,\sigma}{E}\mel{E}{\frac{(z-\frac{\hbar\pi}{T})^2}{(z-\frac{\hbar\pi}{T})^2-E^2} \Gamma[t_0]}{-E}\braket{-E}{j,\sigma}
            \\
        &= \sum_{E = \frac{\hbar\pi}{T}} \frac{1}{2}\expval{\Gamma[t_0]}{E=\frac{\hbar\pi}{T},\text{left edge}}
         \\
        &=\frac{N^{+}_\pi - N^{-}_\pi}{2} =\frac{\pm W_\pi}{2} \,,
    \end{split}
\end{equation}

\end{widetext}
where the sign in the last line is determined by the choice of initial time $t_0$ (see Appendix \ref{app:e}). It is straightforward to conclude that near the vicinity of the poles $z=0$ and $z=\frac{\hbar\pi}{T}$, we have
\begin{equation}
    \lim_{N\to\infty}F(z) \xrightarrow{z\rightarrow E_\omega} \frac{1}{2}\frac{\pm W_\omega}{z-E_\omega}\,.
\end{equation}
This is the one-to-one correspondence between topology and odd-frequency pairing presented in Eq.\,\eqref{SBBC}.
 Since we only relied on chiral symmetry in our proof, the relationship between the accumulated odd-$\omega$ pairing and topology is protected by chiral symmetry and is expected to survive even in the presence of impurities and many-body interaction as we can see in Fig.\ref{fig:disorder}.
 \section{On the sign ambiguity of Eq.\,\eqref{SBBC}} \label{app:e}
 In this appendix, we explain the origin of the sign ambiguity in Eq.\,\eqref{SBBC}. We briefly mentioned in section \ref{section3a} that the existence of chiral symmetry implies that there exist two special initial times $t_0$ and $t_0'$ such that the chiral symmetry operators at these times are equal. Mainly, we have
\begin{equation}\label{eq:chiral_equiv}
    \Gamma[t_0] = \Gamma[t'_0] \equiv \Gamma\,.
\end{equation}
Moreover, the one-period evolution operators at $t_0$ and $t'_0$ are related to each other as follows:
\begin{equation}\label{eq:symmetric time frames}
    U_T[t_0] = \Gamma A^\dagger \Gamma A\,,\qquad\qquad U_T[t'_0] = A\Gamma A^\dagger \Gamma \,,
\end{equation}
 where $A=U(t'_0,t_0)=\mathcal{T}e^{-\frac{i}{\hbar}\int_{t_0}^{t'_0}\dd t H(t)}$ (see \cite{ChiralFloquetWindingNumber}). As we will see next, Eqs.\,\eqref{eq:chiral_equiv} and \eqref{eq:symmetric time frames} explain the sign ambiguity of Eq.\,\eqref{SBBC}.

 The ambiguity of the sign in Eq.\,\eqref{SBBC} originates from the chirality switching of the end modes as one adiabatically changes the initial time from $t_0$ to $t'_0$. To see that, let us consider a topologically protected left edge mode $\ket{\psi_\epsilon[t_0]}$ which is defined at the special initial time $t_0$. It is an eigenstate of the one-period evolution operator $U_T[t_0]$ with eigenvalue $e^{-i\epsilon}$  where  $\epsilon\in\{0,\pi\}$ is the quasienergy in dimensionless units. It is also an eigenstate of the chiral operator: $\Gamma\ket{\psi_\epsilon[t_0]}=e^{-i\gamma}\ket{\psi_\epsilon[t_0]}$ with $\gamma=0, \pi$ corresponding to positive and negative chirality. Now, let us consider the same state but at the other special initial time $t'_0$:   $\ket{\psi_\epsilon[t'_0]}=A\ket{\psi_\epsilon[t_0]}$. We see that
 \begin{equation}\label{eq: chirality switch}
     \begin{split}
         \Gamma\ket{\psi_\epsilon[t'_0]} &= \Gamma A\ket{\psi_\epsilon[t_0]}\\
         &= \Gamma A\Gamma\Gamma\ket{\psi_\epsilon[t_0]}\\
         &= \Gamma A\Gamma e^{-i\gamma}\ket{\psi_\epsilon[t_0]}\\
         &= AA^\dagger\Gamma A\Gamma e^{-i\gamma}\ket{\psi_\epsilon[t_0]} \\
         &= A U_T^\dagger[t_0] e^{-i\gamma}\ket{\psi_\epsilon[t_0]}\\
          &= e^{-i(\gamma-\epsilon)}A\ket{\psi_\epsilon[t_0]} \\
        &= e^{-i(\gamma-\epsilon)}\ket{\psi_\epsilon[t'_0]}\,.
     \end{split}
 \end{equation}
 Eq.\,\eqref{eq: chirality switch} implies that $\ket{\psi_\epsilon[t_0]}$ has the same (opposite) chirality as $\ket{\psi_\epsilon[t'_0]}$ when $\epsilon=0$ $(\epsilon=\pi)$. Because of this chirality switching, we see that the index of the Floquet Hamiltonian $H_{\rm F}[t_0]$ at the gap $z=0$ $(z=\pi\hbar/T)$ should have the same (opposite) sign to the index of $H_{\rm F}[t'_0]$. Since the index is a topological quantity, we expect that the magnitude of the indices at $z=0$ and $z=\pi\hbar/T$ stays constant; however, the sign of the two indices behave differently. As we adiabatically vary the initial time from $t_0$ to $t'_0$, the index at $z=0$ should remain constant while the index at $z=\pi\hbar/T$ will undergo a sign flip. Since the residue of $F(z)$ of Eq.\,\eqref{eq: F_z appndx} at $z=0$ and $z=\pi\hbar/T$ is given by the index, we expect $F(z)$ to display the same behavior. Mainly, we expect the residue of $F(z)$ at $z=0$ should have the same sign as the winding number $W_0$. However, the residue of $F(z)$ at $z=\pi\hbar/T$ can have the opposite sign of the winding number $W_\pi$ depending on the choice of initial time in Eq.\,\eqref{eq: F_z appndx}.
 
\section{Relationship between the effective Green's function and the true Green's function}\label{app:d}
In this appendix, we show how one can obtain the effective Green's function as defined in Eq.\ref{G_def} from the true Green's function associated with the Hamiltonian \ref{eq:Hamiltonian} and vice versa. We also show the equivalence between the true and effective Green's function at stroboscopic times which justifies the validity of our discussion in section \ref{section4} for stroboscopic dynamics. We begin our discussion by recalling that the Green's function for a non-interacting many-body electron system can be defined in terms of the time evolution operator\cite{10.1093/oso/9780198566335.001.0001} as follows:
\begin{equation}\label{eq:green's function alternative definition}
    \mathcal{G}(t,t')=-i\Theta(t-t')U(t,t') =-i\Theta(t-t')\mathcal{T}e^{-i{\frac{1}{\hbar}}\int_{t'}^tH(s)\dd s} \,,
\end{equation}
where $\Theta(t-t')$ is the step function. 
This definition can be easily checked by using the fact that the derivative of the step function is the Dirac delta function and by recalling that the time evolution operator satisfies Schrodinger's equation.

Given the above definition of the Green's function, we can easily manipulate the Green's function using the properties of the time evolution operator. First, let us introduce the Green's function for the Floquet Hamiltonian $H_{\rm F}[t_0]$. According to the definition \eqref{eq:green's function alternative definition}, we see that the Floquet Green's function is given by
\begin{widetext}
\begin{align}
    \mathcal{G}_{\rm F}[t_0](t,t')& = -i\Theta(t-t')e^{\frac{-i(t-t')H_{\rm F}[t_0]}{\hbar}}
    \nonumber
    \\
    &= -i\Theta(t-t')e^{\frac{-i(t-t_0)H_{\rm F}[t_0]}{\hbar}}e^{\frac{i(t'-t_0)H_{\rm F}[t_0]}{\hbar}}
    \nonumber
    \\
    &= -i\Theta(t-t')e^{\frac{-i(t-t_0)H_{\rm F}[t_0]}{\hbar}}U^\dagger(t,t_0)U(t,t_0)U^\dagger(t',t_0)U(t',t_0)e^{\frac{i(t'-t_0)H_{\rm F}[t_0]}{\hbar}}
    \nonumber
    \\
    &= -i\Theta(t-t')P^\dagger(t,t_0)U(t,t_0)U(t_0,t')P(t',t_0)
    \nonumber
    \\
    &= -i\Theta(t-t')P^\dagger(t,t_0)U(t,t')P(t',t_0)
    \nonumber
    \\
    &= P^\dagger(t,t_0)\mathcal{G}(t,t')P(t',t_0)\,,
    \label{eq:Floquet and true green function}
\end{align}
    
\end{widetext}
or, in other words, we have:
\begin{equation}\label{eq:Floquet and true green function2}
    \mathcal{G}(t,t') = P(t,t_0)\mathcal{G}_{\rm F}[t_0](t,t')P^\dagger(t',t_0)\,,
\end{equation}
where $\mathcal{G}(t,t')$ and $\mathcal{G}_{\rm F}[t_0](t,t')$ are the true Green's function of the time-periodic Hamiltonian $H(t)$ and the stroboscopic Green's function of the Floquet Hamiltonian $H_{\rm F}[t_0]$ at initial time $t_0$. Here, we introduced the unitary operator $P(t,t_0)=U(t,t_0)e^{i(t-t_0)H_{\rm F}[t_0]/\hbar}$ otherwise known as the Floquet Kick operator\cite{Floquet.quantum.systems.engineering}.
Eq.\eqref{eq:Floquet and true green function2} defines a mapping between the true Green's function of the time-periodic Hamiltonian $H(t)$ and the stroboscopic Green's function of the Floquet Hamiltonian $H_{\rm F}[t_0]$ at initial time $t_0$. We can verify that it is indeed the correct relation by plugging it into the Schrodinger equation as follows
\begin{widetext}
\begin{align}
    (i&\hbar\partial_t-H(t))\mathcal{G}(t,t') =\Big(i\hbar\partial_t-H(t)\Big)\Big(P(t,t_0)\mathcal{G}_{\rm F}[t_0](t,t')P^\dagger(t',t_0)\Big)
    \nonumber
    \\
    &= \Big(P(t,t_0)i\hbar\partial_t+i\hbar\dot{P}(t,t_0)-H(t)P(t,t_0)\Big)\mathcal{G}_{\rm F}[t_0](t,t')P^\dagger(t',t_0)
    \nonumber
    \\
    &=P(t,t_0)\Big(i\hbar\partial_t-H_{\rm F}[t_0]\Big)\mathcal{G}_{\rm F}[t_0](t,t')P^\dagger(t',t_0)
    \nonumber
    \\
    &= P(t,t_0)\delta(t-t')P^\dagger(t',t_0)
    \nonumber
    \\
    &= \delta(t-t')\,,
\end{align}    
\end{widetext}
Where in the third line, we have used the following identity: 
$$H_{\rm F}[t_0] = P^\dagger(t,t_0)H(t)P(t,t_0)-i\hbar P^\dagger(t,t_0)\dot{P}(t,t_0)$$
We note that Eq.\eqref{eq:Floquet and true green function2} implies that the stroboscopic Green's function is equivalent to the true Green's function at stroboscopic times, i.e. when the difference in the two-time coordinates $t-t'$ is a multiple of the period $T$.
This follows seamlessly from Eq.~\eqref{eq:Floquet and true green function2} and the properties of the operator $P(t,t_0)$. Notice that $P(t,t_0)$ satisfies:
\begin{equation}
    P(nT+t_0,t_0) = P(t_0,t_0) = 1 \qquad\forall n \in \mathbb{Z}\,.
\end{equation}
If we set $t=nT+t_0, t'=n'T+t_0,$ for some arbitrary $n,n'\in\mathbb{Z}$, it follows that
\begin{align}
    \mathcal{G}&(nT+t_0,n'T+t_0) = 
    \nonumber\\
    &P(nT+t_0,t_0)\mathcal{G}_{\rm F}[t_0](nT,n'T)P^\dagger(n'T+t_0,t_0)
    \nonumber\\
    &=\mathcal{G}_{\rm F}[t_0](nT,n'T)\,.
\end{align}

The relationship between the effective Green's function and the true Green's function in the frequency domain is more complicated due to the breaking of the time-translation symmetry. Nevertheless, we can establish a similar relationship to Eq.\eqref{eq:Floquet and true green function2}  in the frequency domain. First, let us recall that the operator $P(t,t_0)$ is periodic in $t-t_0$ with period $T = \frac{2\pi}{\Omega}$. Thus, one can use the following expansion
\begin{equation}\label{eq: kick-operator-expansion}
    P(t,t_0) = \sum_{n=-\infty}^{\infty} P_n(t_0) e^{-i\Omega n(t-t_0)}\,.
\end{equation}
Next, we recall that the Floquet Green's function $\mathcal{G}_{\rm F}[t_0](t,t')$ has time-translation symmetry. Thus, we have $\mathcal{G}_{\rm F}[t_0](t,t') = \mathcal{G}_{\rm F}[t_0](t-t')$. Now, we expand the Floquet Green's function in terms of its Fourier components. We recall that the Floquet Hamiltonian's spectrum is restricted to lie in the region $[-\frac{\hbar\Omega}{2},\frac{\hbar\Omega}{2})$. Thus, the Fourier(frequency) expansion should be given by
\begin{equation}\label{eq: effective-G-expansion}
    \mathcal{G}_{\rm F}[t_0](t-t') = \frac{1}{\hbar}\int_{-\frac{\hbar\Omega}{2}}^{\frac{\hbar\Omega}{2}}e^{-i\omega(t-t')/\hbar}\mathcal{G}_{\rm F}[t_0](\omega)\dd\omega\,.
\end{equation}
 Now, we shift our focus to the true Green's fiction. Let us Fourier transform the true Green's function by integrating the variable $t'$ as follows:
\begin{equation}
    \mathcal{G}(t,\omega) = \int_{-\infty}^{\infty}\dd t' e^{i\omega(t-t')/\hbar}\mathcal{G}(t,t')
\end{equation}
Since the true Green's function is periodic under shifting both time coordinates, it follows that $\mathcal{G}(t,\omega)$ is periodic in the time coordinate $t$. Thus, we apply a discrete Fourier transform.
\begin{equation}
    \mathcal{G}(t,\omega) = \sum_{n=-\infty}^{\infty}\mathcal{G}_n(\omega)e^{-in\Omega t}\,.
\end{equation}
It follows that the true Green's function is given by
\begin{widetext}
    \begin{align}
    \mathcal{G}(t,t') &= \frac{1}{\hbar}\sum_{n=-\infty}^{\infty}\int_{-\infty}^{\infty}\dd\omega\mathcal{G}_n(\omega)e^{-in\Omega t-i\omega(t-t')/\hbar}
    \nonumber
    \\
    &= \frac{1}{\hbar}\sum_{n,m=-\infty}^{\infty}\int_{-\frac{\hbar\Omega}{2}+m\hbar\Omega}^{\frac{\hbar\Omega}{2}+m\hbar\Omega}\dd\omega\mathcal{G}_n(\omega)e^{-in\Omega t-i\omega(t-t')/\hbar}
    \nonumber
    \\
    &= \frac{1}{\hbar}\sum_{n,m=-\infty}^{\infty}\int_{-\frac{\hbar\Omega}{2}}^{\frac{\hbar\Omega}{2}}\dd\omega\mathcal{G}_n(\omega+m\hbar\Omega)e^{-in\Omega t-i(\omega/\hbar+m\Omega)(t-t')}
    \nonumber
    \\
    &= \frac{1}{\hbar}\sum_{n,m=-\infty}^{\infty}\int_{-\frac{\hbar\Omega}{2}}^{\frac{\hbar\Omega}{2}}\dd\omega\mathcal{G}_{n-m}(\omega+m\hbar\Omega)e^{-i(n-m)\Omega t-i(\omega/\hbar+m\Omega)(t-t')}
    \nonumber
    \\
    &= \frac{1}{\hbar}\sum_{n,m=-\infty}^{\infty}\int_{-\frac{\hbar\Omega}{2}}^{\frac{\hbar\Omega}{2}}\dd\omega\mathcal{G}_n(\omega+m\hbar\Omega)e^{-it(\omega/\hbar+n\Omega) + i(\omega/\hbar+m\Omega)t'}\,.
    \label{eq: G-full-expansion}
\end{align}
\end{widetext}
We can identify $\mathcal{G}_{nm}(\omega)=\mathcal{G}_n(\omega+m\Omega)$ which is known as the Green function in the extended Floquet dimension and is more commonly used in literature\cite{high.frequency.Floquet.engineering,Floquet.quantum.systems.engineering}. Using Eqs.~\eqref{eq:Floquet and true green function2}, \eqref{eq: effective-G-expansion}, \eqref{eq: kick-operator-expansion}, and \eqref{eq: G-full-expansion}, we finally arrive at the following identity: 
\begin{equation}\label{eq: Floquet and true green function frequency}
   \mathcal{G}_{nm}(\omega)=P_n(t_0)\mathcal{G}_{\rm F}[t_0](\omega)P_m^\dagger(t_0)\,.
\end{equation}
Thus, we have established a dictionary in Eqs.~\eqref{eq:Floquet and true green function2} and \eqref{eq: Floquet and true green function frequency} that maps the stroboscopic effective Green's function to the full Green's function both in time and frequency domains. Using this dictionary, we can interpret our results in Sect.~\ref{section4}.

\bibliography{biblio}

\end{document}